\documentclass[manuscript,screen]{acmart}

\AtBeginDocument{%
  \providecommand\BibTeX{{%
    \normalfont B\kern-0.5em{\scshape i\kern-0.25em b}\kern-0.8em\TeX}}}

\usepackage{times}
\usepackage{soul}
\usepackage{url}
\usepackage[utf8]{inputenc}
\usepackage{caption}
\usepackage{graphicx}
\usepackage{amsmath}
\usepackage{amsthm}
\usepackage{booktabs}
\usepackage{algorithm}
\usepackage{algorithmic}
\usepackage{mathrsfs}
\usepackage{textpos}

\usepackage{amsmath,amsfonts}
\usepackage{algorithm}
\usepackage{algorithmic}
\usepackage{graphicx}
\usepackage{textcomp}
\usepackage{xcolor}
\usepackage{subfigure}
\usepackage{multirow}
\usepackage{bigstrut}

\newcommand{\liu}[1]{{\color{black}{#1}}}

\newcommand{\TheNameM}{FedDUMAP}
\newcommand{\TheName}{FedDUAP}
\newcommand{\TheAlgoName}{FedDU}
\newcommand{\TheAlgoNameM}{FedDUM}
\newcommand{\ThePruneName}{FedAP}

\newtheorem{assumption}{Assumption}

\newtheorem{theorem}{Theorem}[section]

\newcommand{\jn}[1]{\textcolor{black}{#1}}

\newcommand{\rev}[1]{\textcolor{black}{#1}}

\begin{document}

\title{Efficient Federated Learning Using Dynamic Update and Adaptive Pruning with Momentum on Shared Server Data}

\begin{textblock*}{8cm}(5cm,21cm) 
   {\huge To appear in TIST}
\end{textblock*}

\author{Ji Liu}
\authornote{Corresponding author.}
\authornote{Both authors contributed equally to this research.}
\affiliation{%
  \institution{Hithink RoyalFlush Information Network Co., Ltd.}
  \streetaddress{}
  \city{Hangzhou}
  \country{China}
}

\author{Juncheng Jia}
\authornotemark[2]
\affiliation{%
  \institution{School of Computer Science and Technology, Soochow University and Collaborative Innovation Center of Novel Software Technology and Industrialization}
  \streetaddress{}
  \city{Suzhou}
  \country{China}
}

\author{Hong Zhang}
\authornotemark[2]
\affiliation{%
  \institution{School of Computer Science and Technology, Soochow University}
  \streetaddress{}
  \city{Suzhou}
  \country{China}
}

\author{Yuhui Yun}
\affiliation{%
  \institution{Baidu Research}
  \streetaddress{}
  \city{Beijing}
  \country{China}
}

\author{Leye Wang}
\affiliation{%
  \institution{Key Lab of High Confidence Software Technologies, Ministry of Education, and Software Institute, Peking University}
  \streetaddress{}
  \city{Beijing}
  \country{China}
}

\author{Yang Zhou}
\affiliation{%
  \institution{Department of Computer Science and Software Engineering, Auburn University}
  \streetaddress{}
  \city{Auburn}
  \country{United States}
}

\author{Huaiyu Dai}
\affiliation{%
  \institution{Department of Electrical and Computer Engineering, North Carolina State University}
  \streetaddress{}
  \city{North Carolina}
  \country{United States}
}

\author{Dejing Dou}
\affiliation{%
  \institution{BEDI Cloud and School of Computer Science, Fudan University}
  \streetaddress{}
  \city{Beijing}
  \country{China}
}


\renewcommand{\shortauthors}{Liu et al.}

\begin{abstract}
Despite achieving remarkable performance, Federated Learning (FL) encounters two important problems, i.e., low training efficiency and limited computational resources.
In this paper, we propose a new FL framework, i.e., \jn{\TheNameM{}, with three original contributions,} to leverage the shared insensitive data on the server in addition to the distributed data in edge devices so as to efficiently train a global model.
First, we propose a simple dynamic server update algorithm, which takes advantage of the shared insensitive data on the server while dynamically adjusting the update steps on the server in order to speed up the convergence and improve the accuracy.
\jn{Second, we propose an adaptive optimization method with the dynamic server update algorithm to exploit the global momentum on the server and each local device for superior accuracy.}
Third, we develop a layer-adaptive model pruning method to carry out specific pruning operations, which is adapted to the diverse features of each layer so as to attain an excellent trade-off between effectiveness and efficiency.
Our proposed FL model, \TheNameM{}, combines the three original techniques and has a significantly better performance compared with baseline approaches in terms of efficiency (up to \jn{16.9} times faster), accuracy (up to \jn{20.4}\% higher), and computational cost (up to \jn{62.6}\% smaller).
\end{abstract}


\maketitle

\section{Introduction}

With a large quantity of data distributed in numerous edge devices, 
cloud servers may have some shared insensitive data for efficient processing 
\cite{zhao2018federated}. 
Due to data privacy and security concerns, multiple legal restrictions 
\cite{GDPR,CCPA}
have been put into practice, making it complicated to aggregate the distributed sensitive data within a single high-performance server. As big amounts of training data generally lead to high performance of machine learning models \cite{li2021practical}, \textbf{Federated Learning} (\textbf{FL}) \cite{mcmahan2017communication,liu2022distributed,liu2024enhancing} becomes a promising approach to collaboratively train a model with considerable distributed data while avoiding raw data transfer. 

Traditional FL exploits non-Independent and Identically Distributed (non-IID) data on mobile devices to collaboratively train a global model \cite{mcmahan2017communication}. Within the training process, the weights or the gradients of the global model are transferred between the devices and the server while the raw data stays in each device. FL typically utilizes a parameter server architecture \cite{liu2022distributed,liu2023heterps}, where a parameter server (server) coordinates the distributed training process. The training process generally consists of multiple rounds, each composed of three stages. First, the server chooses a set of devices and broadcasts the global model to them. Second, each of the chosen devices trains the received model using local data and then sends back the updated model. Third, the server gathers the updated models and aggregates them to form a new global model. This process continues when the predefined condition, e.g., a maximum round number or the convergence of the global model, is not achieved.

Although keeping data within the devices can protect data privacy and security, FL encounters two major challenges, which hinder its application in real-world environments. The first challenge is the limited data within a single edge device, which leads to ineffective local training while potential large amounts of shared insensitive data remain useless on the server. The second challenge is the modest edge devices with limited communication and computation capacity \cite{Li2018Learning}, which corresponds to inefficient local training on devices. \rev{To address the first challenge, we leverage the shared insensitive data on the server and design a new FL framework to improve the accuracy of the global model. To address the second challenge, we propose an adaptive optimization method and an adaptive pruning method within the proposed FL framework to improve model accuracy and reduce training cost.}

In addition to the distributed data in edge devices, insensitive data can be transferred to the servers in the cloud to leave space for critical sensitive personal data \cite{Shen2017CloudData,yoshida2020hybrid}, e.g., Amazon Cloud \cite{Amazon}, Microsoft Azure \cite{Azure}, and Baidu AI Cloud \cite{Baidu}.  \rev{The insensitive data may exist on the server by default before the training process. For instance, the server may have prior learning tasks, either training or inference, which have already collected some data to be re-used for the new task of interest. In addition, some end users may offload some insensitive data to the server with certain incentive mechanism such as for crow-sourcing tasks \cite{wang2016toward}, or simply to free up some local storage space \cite{ye2023usability}. Some other recent works are based on the insensitive shared data on the server \cite{yang2022convergence, gu2022fedaux, lian2022blockchain}.} The shared insensitive data can help improve the efficiency of the FL training process \rev{without the restriction of data distribution (see details in Section \ref{subsec:serverUpdate})}.
The shared insensitive data can be transferred to the devices for local training in order to improve the accuracy of the global model \cite{zhao2018federated,YoonSHY21}. 
Although the shared data on the server is not sensitive, they still contain some important personal information. Thus, transferring the shared data to devices may still incur privacy or security problem. 
In addition, this approach leads to significantly high communication overhead.
Furthermore, the existing approaches are inefficient when the shared server data is simply processed as that in edge devices \cite{yoshida2020hybrid} or when knowledge transfer is utilized to deal with heterogeneous models \cite{zhang2021parameterized,lin2020ensemble}.

\jn{Adaptive optimization methods, such as Stochastic Gradient
Descent (SGD) with Momentum (SGDM) \cite{sutskever2013importance}, Adaptive Moment Estimation (Adam), and Adaptive Gradient (AdaGrad), have gained superb advance in accelerating the training speed and the final accuracy. Most of the existing FL literature adopts the adaptive optimization methods either on the server side \cite{duchi2011adaptive,reddi2018adaptive,kingma2014adam,mills2019communication} or on the device side \cite{Yuan2021Federated,mills2019communication,Liu2020Accelerating,gao2021convergence,Wang2020SlowMo,mills2019communication}. However, applying the adaptive optimization on either of them often leads to inefficient learning results with inferior accuracy. The adaptive optimization methods can be exploited on both sides, as shown in \cite{jin2022accelerated}, but the momentum transfer may incur severe communication costs with limited bandwidth between the devices and the server. }

Model pruning reduces the size of a model into a slim one while the accuracy remains acceptable. These kinds of methods can be exploited in the training process to improve the efficiency of the training and to significantly reduce the overhead brought by a big model \cite{jiang2023model}. However, existing pruning approaches incur severe accuracy degradation. The simple application of existing pruning methods in FL does not consider the diverse dimensions and features of each layer \cite{lin2020hrank}. In addition, as a big model may consist of numerous neurons, simple model pruning strategies cannot choose proper portions of the model to prune and may lead to low training efficiency and inferior accuracy \cite{zhao2018federated}.

In this work, we introduce a novel efficient FL framework, i.e., \jn{\textbf{\TheNameM{}}}, which enables collaborative training of a global model based on the sensitive device data and insensitive server data with a powerful server and multiple modest devices. We denote the sensitive data as the data that contain personal information and cannot be shared or transferred according to the users. In addition, the insensitive data can contain personal information while it can be transferred to a cloud server. In order to handle the two aforementioned problems, we utilize the shared insensitive data on the server to improve the accuracy of the global model \jn{with adaptive optimization} with the consideration of the non-IID degrees, which represent the difference between a dataset (either on the server or a device) and the global dataset, i.e., the data on all the devices. 

\TheNameM{} consists of three modules, i.e., \TheAlgoName{}, \TheAlgoNameM{}, and \ThePruneName{}. \TheAlgoName{} is an FL algorithm, which dynamically updates the global model based on the distributed sensitive device data and the shared insensitive server data. In addition, \TheAlgoName{} dynamically adjusts the global model based on the accuracy and the non-IID degrees of the server data and the device data. \jn{Furthermore, we propose a novel adaptive optimization method on top of \TheAlgoName{}, i.e., \TheAlgoNameM{}, to improve the accuracy of the model without transferring the momentum from devices to the server or from the server to the devices. We decouple the optimization on the server and the device sides while exploiting the model generated with adaptive optimization from each device to enable the adaptive optimization on the server side.} Besides, \ThePruneName{} removes useless neurons in the global model to reduce the model size so as to reduce the computational overheads and the computation costs on devices without degrading noticeable accuracy. \ThePruneName{} considers the features of each layer to identify a proper pruning rate, based on the non-IID degrees. Furthermore, \ThePruneName{} prunes the neurons according to the rank values, which can preserve the performance of the model. To the best of our knowledge, we are among the first to propose exploiting non-IID degrees for dynamic global model updates while utilizing adaptive optimization and adaptive pruning for FL. \jn{This manuscript is an extension of a conference version \cite{Zhang2022FedDUAP}.
} In this paper, we make four following contributions:
\begin{enumerate}
    \item A novel dynamic FL algorithm, i.e., \TheAlgoName{}, which utilizes both shared insensitive server data and distributed sensitive device data to collaboratively train a global model. \TheAlgoName{} dynamically updates the global model with the consideration of the model accuracy, normalized gradients from devices, and the non-IID degrees of both the server data and the device data.
    \item \jn{A new adaptive optimization method,
    i.e., \TheAlgoNameM{}, which decouples the optimization between the server side and the device side while exploiting the models generated from devices for the optimization on the server side without additional communication cost.}
    \item An original adaptive pruning method, i.e., \ThePruneName{}, which considers the features of each layer to identify a proper pruning rate based on the non-IID degrees. \ThePruneName{} prunes the model based on the rank values to preserve the performance of the global model.
    \item Extensive experimentation demonstrates significant advantages of \TheNameM{}, including \TheAlgoName{}, \TheAlgoNameM{}, and \ThePruneName{}, in terms of efficiency, accuracy, and computational cost based on three typical models and two real-life datasets.
\end{enumerate}

The rest of this paper is organized as follows. Section \ref{sec:relatedwork} explains the related work. Section \ref{sec:method} proposes our framework, i.e., \jn{\TheNameM{}}, including \TheAlgoName{}, \jn{\TheAlgoNameM{}}, and \ThePruneName{}. Section \ref{sec:experiments} presents the experimental results using three typical models and two real-life datasets. Finally, Section \ref{sec:conclusion} concludes the paper.

\section{Related Work}
\label{sec:relatedwork}

FL was proposed to train a model using the distributed data within multiple devices while only transferring the model or gradients \cite{mcmahan2017communication}. Some works (
\cite{mcmahan2021advances,zhou2022efficient,Li2022FedHiSyn,liu2024aedfl,liu2024fedasmu,che2023fast,liu2023distributed,liu2022multi,che2022federated,liu2023large} and references therein) either focus on the device scheduling or the model aggregation within the server or even with a hierarchical architecture to improve the accuracy of the global model, which only deals with the distributed device data. In order to leave space for sensitive data on devices and with incentive mechanisms \cite{yoshida2020hybrid}, some insensitive data are transferred to the server or the cloud, which can be directly utilized for training. When all the data are transferred to the server, the server data can be considered as IID data \cite{yoshida2020hybrid}. However, transferring all the data including the sensitive data incurs severe privacy and security issues with significant communication costs, which is not realistic \cite{jeong2018communication}.
\rev{Some existing works utilize the insensitive server with heterogeneous models \cite{he2020group}, or within one-shot model training \cite{li2021practical}, based on knowledge transfer methods \cite{dong2022elastic,liu2022elastic}, or label-free data \cite{lin2020ensemble,jeong2020federated}. However, the aforementioned approaches are inefficient \cite{he2020group,li2021practical} or ineffective \cite{lin2020ensemble} without the consideration of the non-IID degrees or big models.} 
Furthermore, while it may lead to significant communication overhead, the transfer of the insensitive server data to devices \cite{zhao2018federated,YoonSHY21} may incur severe privacy and security issues. 
The proper devices can be selected for training based on the server data \cite{nagalapatti2021game}, which can be integrated with our proposed approach.

\jn{Without adaptive optimization within the model aggregation process, traditional methods, e.g., FedAvg \cite{mcmahan2017communication}, may incur the client drift problem, which makes the global model over-fitted to local device data \cite{karimireddy2020scaffold}. While control parameters \cite{karimireddy2020scaffold} can help alleviate the problem, they require the devices to participate all through the training process \cite{karimireddy2020scaffold,acar2021federated}, which may not be feasible in cross-device FL. Adaptive optimization methods, e.g., AdaGrad \cite{duchi2011adaptive}, Yogi \cite{reddi2018adaptive}, Adam \cite{kingma2014adam,mills2019communication} and momentum \cite{rothchild2020fetchsgd,mills2021accelerating,che2023federated,jin2022accelerated} can be exploited to address the client drift problem. However, the existing approaches generally consider only one side, i.e., either server side \cite{reddi2021adaptive} or device side \cite{Yuan2021Federated}, which may lead to inferior accuracy. 
Furthermore, the devices may have heterogeneous non-IID data, which makes the direct application of the momentum method insufficient \cite{Liu2020Accelerating}. Although adaptive optimization is applied on both the device side and the server side in some recent works \cite{jin2022accelerated}, the communication of momentum between the server and devices may incur high communication costs.}

\begin{figure}[!t]
\centering
\includegraphics[width=0.8\linewidth]{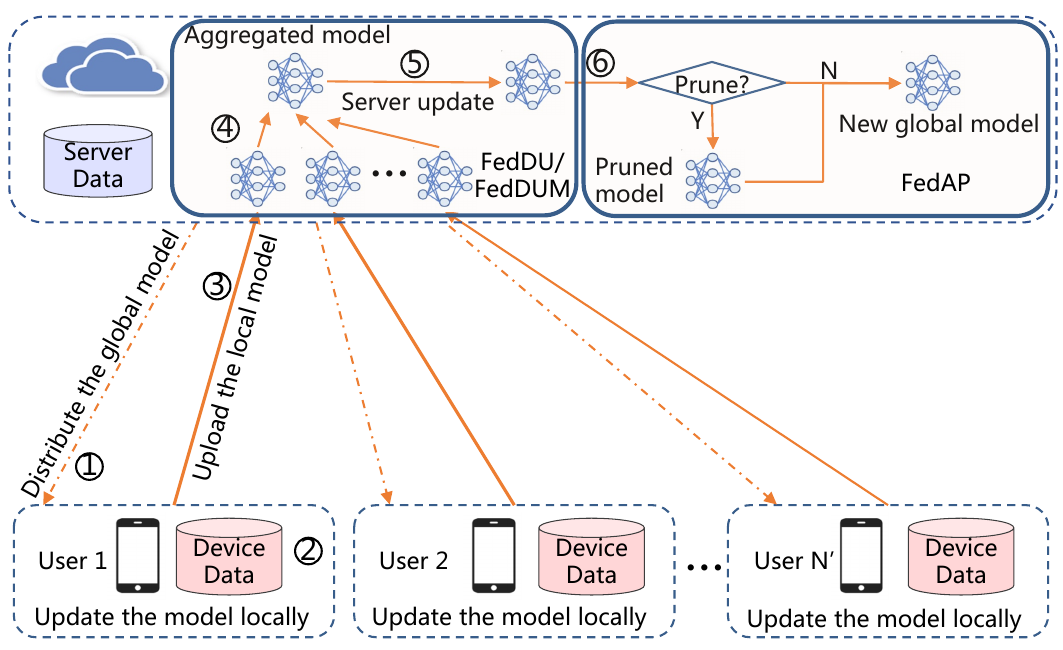}
\caption{The training process of \TheNameM{} Framework.}
\label{fig:framework}
\end{figure}

\rev{Model pruning can reduce the size of the model, which corresponds to smaller communication overhead and computation costs compared with the original model within the training process \cite{jiang2023model, li2021hermes, li2021fedmask, deng2022tailorfl} and the inference process \cite{yu2021adaptive} of FL.} However, the shared insensitive server data is seldom utilized. Two types of techniques exist for the pruning process, including filter (structured) pruning and weight (unstructured) pruning. The unstructured pruning set the chosen parameters to 0 while the structure of the original model remains unchanged. Although the accuracy remains almost the same as that of the original one \cite{zhang2021validating} while reducing communication overhead \cite{jiang2023model}, the unstructured pruning cannot reduce computational cost with general-purpose computing resources, e.g., edge devices \cite{lin2020hrank}. On contrary, structured pruning can modify the model structure by abandoning selected neurons or filters in the original model. As some parts of the original model are removed, the structured pruning leads to significantly smaller computational costs and communication overhead \cite{lin2020hrank}. However, it is non-trivial to determine the pruning rate, which hinders the application of unstructured pruning in the training process of FL. Sparsification, quantization \cite{konevcny2016federated,jia2024efficient}, or dropout \cite{Bouacida2021FedDropout,chen2022fedobd} are efficient techniques to reduce the communication overhead as well, which is out of the scope of this paper and can be combined with our proposed approach for better performance.

\jn{In this paper, we utilize both the shared insensitive server data and the distributed sensitive device data for FL training. We propose an FL framework, i.e., \TheNameM{}, consisting of \TheAlgoName{}, \TheAlgoNameM{}, and \ThePruneName{}. \TheAlgoName{} dynamically adjusts the global model with the shared insensitive server data based on the non-IID degrees and model accuracy. \TheAlgoNameM{} considers an adaptive optimization method on both the device and the server sides without additional communication overhead. Furthermore, \ThePruneName{} prunes the global model with the consideration of the features of each layer and removes the filters based on the rank values to reduce communication and computational costs while achieving high accuracy. The pruning rate is calculated based on the non-IID degrees. \TheNameM{} can achieve high accuracy and efficient training at the same time.}

\section{Method}
\label{sec:method}

In this section, we propose our FL framework, i.e., \TheNameM{}. We first present the system model. Then, we detail \TheAlgoName{}, which dynamically updates the global model. Afterward, we explain \jn{the adaptive momentum-based optimization, i.e., \TheAlgoNameM{}}, to improve the accuracy of the global model. Finally, we reveal the design of \ThePruneName{}, which reduces the size of the global model to reduce communication and computational costs. 

\subsection{System Model}
\label{subsec:systemModel}

\begin{table*}[ht]
\small
\caption{Summary of main notations}
\begin{center}
\resizebox{\textwidth}{!}{
\begin{tabular}{cl}
\hline
Notation & Definition \\
\hline
$N$; $N'$ & The number of edge devices; the number of devices selected for an iteration; the number of selected devices \\
$D_k$; $D_0$; $k$ & The local dataset on Device $k$; the server data; the index of a device\\
$n_k$; $n_0$ & The number of samples on Device $k$; the number of samples on the server\\
$x_{k, j}$; $y_{k, j}$ & The $j$-th input data sample in Device $k$; the label of $x_{k, j}$ \\
$\mathcal{X}$; $\mathcal{Y}$ & The data sample set; the label set \\
$w$; $w^{t}$; $w^{t - \frac{1}{2}}$ & The parameters of the global model; the global model in Round $t$; the aggregated model in Round $t$ \\
$P_k = \{P_k(y)|y\in \mathcal{Y}\}$ & The distribution of data in Device $k$ with 0 representing the server \\
$P_k(y)$ & The probability that a data sample corresponds to Label $y$ in Device $k$ \\
$P_m$; $\overline{P}$ & The average distribution of two datasets; the distribution of all the device data\\
$\overline{P'}^t$ & The distribution of all the data in the selected devices in Round $t$\\
$\mathcal{D}(P_k)$ & The non-IID degree of data on Device $k$\\
$\mathscr{D}$; $\mathscr{D}^t$ & The set of all devices and the server; the set of selected devices in Round $t$ \\
$F_k(w)$ & The local loss function in Device $k$\\
$E$; $B$; $\eta$ &  The local epoch; the local batch size; the learning rate\\
$\tau$; $\tau_{eff}^{t}$ & The number of iterations performed in the server update; the effective steps for the server update in Round $t$ \\
$\overline{g}_0^{t}(\cdot)$ &  The normalized stochastic gradients on the server in Round $t$ \\
$n'$; $n_0$ & The total number of samples in all the selected devices; the number of samples in the server data \\
$w^{t - \frac{1}{2}, i}$ & The parameters of the $i$-th local iteration of the updated aggregated model in Round $t$\\
$acc^t$ & The accuracy of the aggregated model in Round $t$\\
$W_k$; $W'_k$ & The initial parameters of the local model in Device $k$; the parameters of the local model in Device $k$ after $T$ rounds\\
$H(W'_k)$ & The Hessian matrix of the loss function on Device $k$ \\
$\lambda_k^m$; $d_k$ & The $m$-th eigenvalue in the Hessian matrix on Device $k$; the rank of the Hessian matrix on Device $k$ \\
$p^*$; $p^*_k$; $p^*_l$ & The expected global pruning rate; the expected pruning rate on Device $k$; the expected pruning rate of Layer $l$ \\
$\epsilon$ & A small value to avoid the division by 0 \\
$\mathscr{M}$; $q_l$ & The number of parameters in the model; the number of parameters in Layer $l$ \\
$r^j_l$; $d_l$ & The rank of the $j$-th filter in Layer $l$; the number of filters in Layer $l$ \\
\hline
\end{tabular}
}
\end{center}
\label{tab:notation}
\end{table*}

Figure \ref{fig:framework} shows the training process of \TheNameM{}, where the FL system consists of a server and $N$ edge devices. The description of major notations is summarized in Table \ref{tab:notation}.
We consider a powerful server and modest devices. The shared insensitive data is stored on the server, and the distributed sensitive data is dispersed in edge devices. 
Both the server data and the device data can be utilized to update the model during the FL training process of FL.
Let us assume that a dataset $D_k = \{x_{k,j}, y_{k, j}\}_{j=1}^{n_k}$, composed of $n_k$ samples, resides on Device $k$. We take $D_0$ as the server data, and $D_k, k\neq 0$ as the device data. $x_{k,j}$ refers to the $j$-th sample on Device $k$, while $y_{k,j}$ represents the corresponding label. We utilize $\mathcal{X}$ to represent the set of samples and $\mathcal{Y}$ to represent the set of labels.
Then, we formulate the objective of the training process as follows:
\begin{equation}
\min_{w}F(w)\textrm{, with }F(w)\triangleq\frac{1}{n}\sum_{k = 1}^N n_k F_k(w),
\label{eq:problem}
\end{equation}
where $F_k(w)\triangleq\frac{1}{n_k}\sum_{\{x_{k,j},y_{k,j}\} \in \mathcal{D}_k} f(w,x_{k,j},y_{k,j})$ represents the loss function on Device $k$ with $f(w,x_{k,j},y_{k,j})$ capturing the error of the inference results based on the model with the sample pair $\{x_{k,j},y_{k,j}\}$ and $w$ refers to the global model.

With the distributed non-IID data in FL \cite{mcmahan2021advances}, we utilize the Jensen–Shannon (JS) divergence~\cite{fuglede2004jensen} to quantize the non-IID degree of a dataset as shown in Formula \ref{eq:js-divergence}:
\begin{equation}
\mathcal{D}(P_k) = \frac{1}{2}\mathcal{D}_{KL}(P_k||P_m) + \frac{1}{2}\mathcal{D}_{KL}(\overline{P}||P_m),
\label{eq:js-divergence}
\end{equation}
where $P_m = \frac{1}{2}(P_k + \overline{P})$, $\overline{P} =\frac{\sum_{k = 1}^{N}n_k P_k}{\sum_{k = 1}^{N}n_k}$, $P_k = \{P_k(y)|y \in \mathcal{Y}\}$, $P_k(y)$ refers to the possibility that a sample is related to Label $y$, and $\mathcal{D}_{KL}(\cdot||\cdot)$ corresponds to the Kullback-Leibler (KL) divergence~\cite{kullback1997information} as defined in Formula \ref{eq:kl-divergence}: 
\begin{equation}
\mathcal{D}_{KL}(P_i||P_j)= \sum_{y \in \mathcal{Y}}P_i(y)\log\left(\frac{P_i(y)}{P_j(y)}\right).
\label{eq:kl-divergence}
\end{equation}
When the distribution of a dataset differs from that of the global dataset more significantly, the corresponding non-IID degree becomes higher. Although the raw data is not transferred from devices to the server or from the server to the devices in \TheNameM{}, we assume that $P_k$ and $n_k$ incur little privacy concern \cite{lai2021oort}, and can be transferred from devices to the server before the FL training process. When this statistical meta information, i.e., $P_k(y)$, cannot be shared because of restrictions, we can utilize gradients to calculate the data distribution of the dataset on each device \cite{yang2022improved}.

The FL training process contains multiple rounds in \TheNameM{}, each of which consists of 6 steps. In Step \textcircled{1}, a set of devices ($\mathscr{D}^t$ with $t$ referring to the round number) is randomly selected to participate in the training process of Round $t$ and the server broadcasts the global model to the device of $\mathscr{D}^t$. Afterward, each device updates the received model with the adaptive momentum-based optimization based on the local data in Step \textcircled{2}. Then, in Step \textcircled{3}, the selected devices upload updated models to the server, and the server aggregates all the received models based on FedAvg \cite{mcmahan2017communication} in Step \textcircled{4}. In addition, the server updates (see details of the server update in Section \ref{subsec:serverUpdate}) aggregated model utilizing the shared insensitive server data and \jn{the adaptive momentum-based optimization method (see details in Section \ref{subsec:optimization})} in Step \textcircled{5}. Furthermore, in a predefined round, the server performs model pruning based on the accuracy of the global model and the non-IID degrees (see details in Section \ref{subsec:pruning}). While Steps \textcircled{1} - \textcircled{4} resemble those within traditional FL, we propose exploiting the insensitive server data to improve the performance of the global model (in FedDU) with adaptive momentum-based optimization (in FedDUM) on both the server and device sides (Step \textcircled{5}) and an adaptive model pruning method (in FedAP) to improve the efficiency of the training process (Step \textcircled{6}).

\subsection{Server Update}
\label{subsec:serverUpdate}

In this section, we explain the design of \TheAlgoName{}, which exploit both the shared insensitive server data and the distributed sensitive device data to adjust the global model. We exploit the non-IID degrees and the model accuracy to determine the weights of aggregated model and those of the normalized gradients based on the server data in \TheAlgoName{}, so as to avoid over-fitting to the server data. 

We assume that the size of the insensitive server data is significantly bigger than the dataset within a single device. In this case, the straightforward combination of the aggregated model from the devices and the  gradients calculated based on the shared insensitive server data leads to inferior accuracy due to objective inconsistency \cite{wang2020tackling}. Inspired by \cite{wang2020tackling}, we normalize the gradients generated using the server data to deal with this issue. The model aggregation in \TheAlgoName{} is formulated as Formula \ref{eq:FedGSS}. 
\begin{equation}
w^t = w^{t - \frac{1}{2}} - \tau_{eff}^{t - 1}\eta \overline{g}_0^{(t - 1)}(w^{t - \frac{1}{2}}),
\label{eq:FedGSS}
\end{equation}
where $w^t$ refers to the weights of the global model in Round $t$, $w^{t - \frac{1}{2}}$ is the weights of the aggregated model from the selected devices in Round $t$ as formulated in Formula \ref{eq:FedGSS-device} \cite{mcmahan2017communication}, $\tau_{eff}^{t - 1}$ corresponds to the size of effective steps for normalized gradients calculated based on the shared insensitive server data as shown in Formula \ref{eq:effectiveStep}, $\eta$ represents the learning rate, $\overline{g}_0^{t}(\cdot)$ refers to the normalized gradients generated based on the server data in Round $t$ as formulated in Formula \ref{eq:normalization}.
\begin{equation}
w^{t - \frac{1}{2}} = \sum_{k \in \mathscr{D}^t}\frac{n_k}{n'} (w^{t - 1} - \eta g_k^{t - 1}(w^{t - 1})),
\label{eq:FedGSS-device}
\end{equation}
where $n' = \sum_{k \in \mathscr{D}^t}n_k$ refers to the size of the dataset on all the selected devices, $g_k^{t-1}(\cdot)$ corresponds to the gradients generated using the dataset on Device $k$.
\begin{equation}
\overline{g}_0^{(t-1)}(w^{t - \frac{1}{2}})=\frac{\sum_{i=1}^{\tau}g_0^{(t-1)}(w^{t-\frac{1}{2},i})}{\tau},
\label{eq:normalization}
\end{equation}
where $w^{t-\frac{1}{2},i}$ represents the parameters of the global model after aggregating the gradients based on the server data, at $i-$th iteration of Round $t$, $g_0^{(t-1)}(\cdot)$ refers to the gradients calculated using the server data, and $\tau =\lceil \frac{|n_0| E}{B} \rceil$ corresponds to the number of iterations of Round $t$, $E$ is the number of local epochs, $B$ is the batch size. Each round corresponds to multiple iterations with a mini-batch of sampled shared insensitive server data.

While $\tau_{eff}^{t}$ has a significant impact on the training process, we dynamically choose an effective step size with the consideration of the model accuracy, the non-IID degrees of both the server data and the device data, with the number of rounds, as formulated in Formula \ref{eq:effectiveStep}.
\begin{equation}
\tau_{eff}^{t} = f'(acc^t) * \frac{n_0 * \mathcal{D}(\overline{P'}^t)}{n_0 * \mathcal{D}(\overline{P'}^t) + n' * \mathcal{D}(P_0)} * \mathcal{C} * decay^t * \tau,
\label{eq:effectiveStep}
\end{equation}
where $acc^t$ represents the model accuracy calculated based on the server data in Round $t$, i.e., $w^{t - \frac{1}{2}}$ defined in Formula \ref{eq:FedGSS-device}, $n_0$ corresponds to the size of the shared insensitive server data, $\mathcal{D}(\cdot)$ is shown in Formula \ref{eq:js-divergence}, $\overline{P'}^t = \frac{\sum_{k \in \mathscr{D}^t}n_k P_k}{\sum_{k \in \mathscr{D}^t}n_k}$ refers to the distribution of the distributed sensitive data in all the chosen devices in Round $t$, $P_0$ is the distribution of the shared insensitive server data, $decay \in (0, 1)$ is utilized to ensure the convergence of the global model towards the solution of Formula \ref{eq:problem}, and $\mathcal{C}$ corresponds to a hyper-parameter.
$f'(acc)$ is calculated using $acc$.
$acc$ is small at the beginning of the training. In this stage, the value of $f'(acc)$ should be prominent so as to take advantage of the server data to improve the accuracy of the global model. Then, at a late stage of the training process, $f'(acc)$ should be small to attain the objective defined in Formula \ref{eq:problem} while avoiding over-fitting to the shared insensitive server data. 

When the distribution of server data resembles that of the overall device data, i.e., $\mathcal{D}(P_0)$ is small, or the distribution of the data of the chosen devices differs much from the overall device data, i.e., $\mathcal{D}(\overline{P'}^t)$ is significant, we take a significant value of $\tau_{eff}^{t}$ to improve the accuracy of the global model. 
Furthermore, the data size improves the importance of the dataset, as well. In addition, the weight of the device data is $\frac{n'}{\mathcal{D}(\overline{P'}^t)}$ and that of the server data is $\frac{n_0}{\mathcal{D}(P_0)}$. Finally, we have $\frac{\frac{n_0}{\mathcal{D}(P_0)}}{\frac{n_0}{\mathcal{D}(P_0)} + \frac{n'}{\mathcal{D}(\overline{P'}^t)}}$ as the importance of the server data, which is the second part of Formula \ref{eq:effectiveStep}.

\begin{figure}[t]
\begin{algorithm}[H]
\caption{Federated Dynamic Server Update (\TheAlgoName{})}
\label{alg:dyamic}
\begin{algorithmic}[1]
\REQUIRE  \quad \\
$\mathscr{D}^t$: The set of selected devices in Round $t$ \\
$\mathcal{D}_k$: The dataset on Device $k$ with 0 representing that on the server \\ 
$w^{t-1}$: The global model in Round $t-1$ \\ 
$E$: The number of local epochs \\
$B$: The local batch size \\
$decay$: The decay rate \\
$P$: The set of data distribution $\{P_k|k \in \{0\} \cup \mathscr{D}^t\}$ with 0 representing the server \\ 
$\eta$: The learning rate
\ENSURE \quad \\
$w^{t}$: The global model in Round $t$ 
\FOR{$k$ in $\mathscr{D}^t$ (in parallel)} \label{line:localUpdateBegin}
\STATE Calculate $g_k^{t - 1}(w^{t - 1})$ using $w^{t-1}$,  $\mathcal{D}_k$
\ENDFOR \label{line:localUpdateEnd}
\STATE Calculate $w^{t - \frac{1}{2}}$ using $w^{t-1}, \eta$ according to Formula \ref{eq:FedGSS-device} \label{line:aggregatedModel}
\STATE Calculate $w^{t}$ using $w^{t-\frac{1}{2}}, decay, E, B, P, \eta$ according to Formula \ref{eq:FedGSS} \label{line:serverUpdate}
\end{algorithmic}
\end{algorithm}
\end{figure}

Algorithm \ref{alg:dyamic} explains \TheAlgoName{}. The model is updated using the local dataset on each chosen device (Lines \ref{line:localUpdateBegin} - \ref{line:localUpdateEnd}). Afterward, the models are aggregated using Formula \ref{eq:FedGSS-device} (Line \ref{line:aggregatedModel}). Finally, the aggregated model is updated utilizing the shared insensitive server data based on Formula \ref{eq:FedGSS} (Line \ref{line:serverUpdate}).

\jn{Let us assume that the expected squared norm of stochastic gradients on the server is bounded, i.e., $\mathop{\mathbb{E}}||\overline{g}_0||^2 \leq G^2$. Then, in Round $t$, $\tau_{eff}^t \leq C \cdot decay^t \cdot \tau$. Afterward, the server update term can be less than  $ C \cdot decay^t \cdot \tau \cdot \eta \cdot G$. In this case, after sufficient steps, the server update becomes negligible, i.e., $\mathop{\mathbb{E}}[\lim_{t \to \infty} \tau_{eff}^{t-1} \eta \overline{g}_0^{t-1}(w^{t-\frac{1}{2}})] \leq \lim_{t \to \infty} C \cdot decay^{t-1} \cdot \tau \cdot \eta \cdot G = 0$, with $0 < decay < 1$. Finally, \TheAlgoName{} degrades to FedAvg \cite{mcmahan2017communication}, the convergence of which is guaranteed with a decaying learning rate $\eta$  \cite{Li2020On,Zhou2018On}.}

\subsection{\jn{Momentum-based Optimization}}
\label{subsec:optimization}

\jn{In order to further improve the accuracy of the global model, we exploit momentum within the server update on the server and within the local iteration on each device. In this section, we propose a simple adaptive momentum approach, i.e., \TheAlgoNameM{}, which enables the optimization on both the server and the devices without additional communication cost.}

\jn{The centralized SGDM can be formulated as:
\begin{equation}\label{eq:sgdm}
\begin{split}
m^t  = \beta * m^{t - 1} + (1 - \beta) *g(w^{t - 1}), w^t = w^{t - 1} - \eta * m^t,
\end{split}
\end{equation}
where $m$ is momentum. A simple application of the momentum into the FL environment is to decompose the SGDM to each device while aggregating the momentum as formulated in Formula \ref{eq:fedda}.
\begin{equation}
\label{eq:fedda}
\begin{split}
&m^t_k  = \beta * m^{t - 1}_k + (1 - \beta) *g_k(w_k^{t - 1}),\\
&w_k^t = w_k^{t - 1} - \eta * m_k^t,\\
&m^t = \sum_{k \in \mathscr{D}^t} \frac{n_{k}}{n'}m_k^{t - 1},\\ 
&w^t = \sum_{k \in \mathscr{D}^t} \frac{n_{k}}{n'}w_k^{t - 1}. 
\end{split}
\end{equation}
In order to extend the optimization into multiple local epochs, we can formulate the process as follows:
\begin{equation}\label{Naive}
\begin{split}
&{m'}_k^{t, t'}  = \beta' * {m'}_k^{t, t' - 1} + (1 - \beta') * {g'}_k({w'}_k^{t, t' - 1}),\\
&{w'}_k^{t, t'} = {w'}_k^{t, t' - 1} - \eta' * {m'}_k^{t, t'},\\
&m^t = \sum_{k \in \mathscr{D}^t} \frac{n_k}{n'}{m'}_k^{t, E - 1},\\ 
&w^t = \sum_{k \in \mathscr{D}^t} \frac{n_k}{n'}{w'}_k^{t, E - 1},
\end{split}
\end{equation}
\begin{figure}[!t]
\begin{algorithm}[H]
\caption{Federated Dynamic Update with Momentum (\TheAlgoNameM{})}
\label{alg:dyamicM}
\begin{algorithmic}[1]
\REQUIRE  \quad \\
$\mathscr{D}^t$: The set of selected devices in Round $t$ \\
$\mathcal{D}_k$: The dataset on Device $k$ with 0 representing that on the server \\ 
$w^{t-1}$: The global model in Round $t-1$ \\ 
$E$: The number of local epochs \\
$B$: The local batch size \\
$decay$: The decay rate \\
$P$: The set of data distribution $\{P_k|k \in \{0\} \cup \mathscr{D}^t\}$ with 0 representing the server \\ 
$\eta$: The learning rate
\ENSURE \quad \\
$w^{t}$: The global model in Round $t$ 
\FOR{$k$ in $\mathscr{D}^t$ (in parallel)} \label{line:localMUpdateBegin}
\STATE Calculate ${w'}_k^{t - 1, E - 1}$ using $w^{t-1}$,  $\mathcal{D}_k$, $B$, $E$ using Formula \ref{eq:simpleMD}
\ENDFOR \label{line:localMUpdateEnd}
\STATE Calculate $g(w^{t - 1})$ using ${w'}_k^{t - 1, E - 1}, decay, E, B, P, \eta$ according to Formula \ref{eq:simpleMS} \label{line:calculateGradient}
\STATE Calculate $w^{t}$ using $g(w^{t - 1})$ according to Formula \ref{eq:sgdm} \label{line:serverUpdateM}
\end{algorithmic}
\end{algorithm}
\end{figure}
where ${m'}_k^{t, t'}$ represents the local momentum within Device $k$ at global Round $t$ and local Iteration $t'$, $\beta'$ is the local weight for gradients, ${g'}_k$ refers to the gradients with the parameters of the model ${w'}_k^{t, t'}$, and $\eta'$ is the local learning rate. Please note that ${g'}_k$ and $\eta'$ are local hyper-parameters on each device and are different from those in Formulas \ref{eq:sgdm} and \ref{eq:fedda}, which refer to global parameters on the server. We have the first momentum equal to the global momentum, i.e., ${m'}_k^{t, 0} = m^t$, and the first parameters of the model equal to the global parameters, i.e., ${w'}_k^{t, 0} = w^t$. However, this simple decomposition may deviate from the momentum of the centralized optimizer and harm the local training \cite{jin2022accelerated}. In addition, the communication of momentum, i.e., ${m'}_k^{t, E - 1}$ and $m^t$, may incur significant communication cost. Inspired by \cite{wang2021local}, we restart the local adaptive optimization and set the momentum buffers to zero at the beginning of local iterations. Furthermore, we calculate the difference between the updated model and the original model as the total gradients as defined in Formula \ref{eq:simpleMS}. Then, we can utilize Formula \ref{eq:simpleMD} below to perform the local iteration in each selected device, which replaces Formula \ref{eq:FedGSS-device} in Algorithm \ref{alg:dyamic} with adaptive momentum:
\begin{equation}\label{eq:simpleMD}
\begin{split}
{m'}_k^{t, t'}  = \beta' * {m'}_k^{t, t' - 1} + (1 - \beta') * {g'}_k({w'}_k^{t, t' - 1}), {w'}_k^{t, t'} = {w'}_k^{t, t' - 1} - \eta' * {m'}_k^{t, t'},
\end{split}
\end{equation}
with ${m'}_k^{t, 0} = 0$ and ${w'}_k^{t, 0} = w^t$. This operation only requires the parameters of the model transferred from the server to devices without additional communication costs compared with FedAvg. Afterward, we exploit Formula \ref{eq:simpleMS} below to calculate the gradients on the server, and utilize Formula \ref{eq:sgdm} to replace Formula \ref{eq:FedGSS} in Algorithm \ref{alg:dyamic} with adaptive momentum:
\begin{equation}\label{eq:simpleMS}
\begin{split}
w^{t - \frac{1}{2}} = \sum_{k \in \mathscr{D}^t} \frac{n_k}{n'}{w'}_k^{t - 1, E - 1}, g(w^{t - 1})  = w^{t - \frac{1}{2}} + \tau_{eff}^{t - 1}\eta \overline{g}_0^{(t - 1)}(w^{t - \frac{1}{2}}) - w^t,
\end{split}
\end{equation}
where $\tau_{eff}^{t - 1}\eta \overline{g}_0^{(t - 1)}(w^{t - \frac{1}{2}})$ is the same as that in Formula \ref{eq:FedGSS}. This only requires the transfer of the parameters of the local updated model, which does not incur additional communication cost compared with \TheAlgoName{}. }

\jn{\TheAlgoNameM{} is shown in Algorithm \ref{alg:dyamicM}. The local model update is performed according to Formula \ref{eq:simpleMD} in parallel in each selected device (Lines \ref{line:localMUpdateBegin} - \ref{line:localMUpdateEnd}). Then, the aggregated gradient is calculated using Formula \ref{eq:simpleMS} (Line \ref{line:calculateGradient}). Afterward, the new global model is updated using the gradients and momentum according to Formula \ref{eq:sgdm} (Line \ref{line:serverUpdateM}).}

\begin{figure}[!t]
\begin{algorithm}[H]
\caption{Federated Adaptive Structured Pruning (\ThePruneName{})}
\label{alg:automatic_pruning}
\begin{algorithmic}[1]
\REQUIRE \quad \\
$L$: The list of convolutional layers to prune\\
$\mathscr{D}$: The set of all devices and the server \\
$w$: The initial model\\
$w^*$: The current model in Round $t$\\
$W = [v_1,v_2,\cdots,v_\mathscr{M}]$: The list of parameters in Model $w^*$ 
\ENSURE \quad \\
$w'$: The pruned model in Round $t$
\STATE $w' \leftarrow w^*$
\FOR{$k \in \mathscr{D}$ (in parallel)}  \label{line:localPRateBegin} 
        \STATE Calculate the expected pruning rate $p^*_{k}$ \label{line:localPRate}
    \ENDFOR \label{line:localPRateEnd}
\STATE Calculate $p^*$ according to Formula \ref{eq:FedASP} \label{line:fedPRateEnd}
\label{line:eachPRateBegin} 
\STATE $W=[v_{o_1}, v_{o_2}, \cdots, v_{o_R}] \leftarrow$ Sort $W$ in ascending order of $|v|$ \label{line:globalSort}
\STATE $\mathscr{V} = |v_{o_{\lfloor R*p^* \rfloor}}|$ \label{line:threshold}
\FOR{$l \in L$} \label{line:cnnBegin}
    \STATE $W_l = [v_1, v_2,\cdots,v_{q_l}]\leftarrow$ The parameters in Layer $l$ \label{line:localSort}
    \STATE $W'_l \leftarrow [v_1, v_2, \cdots, v_{q'_l}]$ with each $|v_q| < \mathscr{V}$ \label{line:localSelect}
    \STATE $p_l^*\leftarrow \frac{Cardinality(W'_l)}{q_l}$ \label{line:eachPRateEnd} 
    \STATE Calculate the ranks $R_l$ of each filter in Layer $l$ \label{line:rank}
    \STATE Sort the filters according to $R_l$ in an ascending order \label{line:sort}
    \STATE $w'_l \leftarrow$ Preserve the last $d_l - \lfloor p^*_l * d_l \rfloor$ filters in $R_l$ \label{line:prune}
    \STATE $w' \leftarrow$ Replace the $l$-th layer of $w'$ with $w'_l$ \label{line:replace}
\ENDFOR \label{line:cnnEnd}
\end{algorithmic}
\end{algorithm}
\end{figure}

\jn{In order to analyze the adaptive optimization on devices, we have the following assumptions:
\begin{assumption} 
\label{assump:lip}
(Lipschitz Gradient). There exists a constant $L_g$ such that $\|g_k(w_1) - g_k(w_2)\|\le L_g\|w_1-w_2\|$ for any $w_1, w_2$ and $k = 1,\cdots,N$.
\end{assumption}
\begin{assumption} 
\label{assump:boundg}
(Bounded Gradient) $\|g_k\|_{L^\infty}<\infty$ for any $k=1,2,\cdots,N$.
\end{assumption}
\begin{assumption} 
\label{assump:boundm}
(Bounded Momentum) $\|m_k\|_{L^\infty}<\infty$ for any $k=1,2,\cdots,N$.
\end{assumption}
Then, we can get the following theorem:
\begin{theorem}
Local momentum deviates from the centralized one at linear rate $O(e^{\lambda^+E})$.
\end{theorem}}

\begin{proof}
\jn{From the perspective of dynamics of Ordinary Differential Equations (ODE)  \cite{jin2022accelerated}, we can formulate the difference between the local optimization and the centralized optimization as:}

\begin{equation}\label{diffmatrix}
\begin{split}
\begin{pmatrix}\|m_k^{t, E-1} - m^{t, E-1}\| \\ \|w_k^{t, E-1} - w^{t, E-1}\| \end{pmatrix} \le e^{AE}\begin{pmatrix}\|m_k^{t,0} - m^t\| \\ \|w_k^{t,0} - w^t\| \end{pmatrix} + (1-\beta')\int_0^{E-1} e^{A(E-t')}\begin{pmatrix}\|R^{t, t'}\| /\eta'\\ 0 \end{pmatrix}dt'.
\end{split}
\end{equation}
\jn{where $m^{t, t'}$ represents the momentum in centralized optimization, $w^{t, t'}$ represents the model in centralized optimization, $R^{t, t'} = \sum_{j \in \mathscr{D}^t, j\ne k}\frac{n_j}{n'}\big(g_j(w^{t, t'}) - g_k(w^{t, t'})\big)$, and $ A = \begin{pmatrix} -(1-\beta')/\eta'&(1-\beta')L_g/\eta' \\ 1& 0 \end{pmatrix}$. The eigenvalue of Matrix $A$ is $\lambda^\pm = \frac{-(1-\beta')\pm \sqrt{(1-\beta')^2+4(1-\beta')L_g}}{2\eta'}$. Note that $m_k^{t,0} = 0$ and $w_k^{t,0} = w^t$. In addition, $\|e^{At}\|\le C_Ae^{\lambda^+ t'}$ for any $t'\ge 0$ for some constant $C_A$. Then, we have:}
\begin{equation}
\begin{split}
&\|m_k^{t, t'} - m^{t, t'}\| + \|w_k^{t, t'} - w^{t, t'}\|\\
\le~&C_A\|m^t\|e^{\lambda^+ t'} + e^{\lambda^+t'}(1-\beta')\int_0^{t'} e^{-\lambda^+ \tau}\begin{pmatrix}\|R^{t, t'}\| /\eta'\end{pmatrix}d\tau\\
\le~&C_A\|m^t\|e^{\lambda^+ E} + e^{\lambda^+E}(1-\beta')\int_0^{E-1} e^{-\lambda^+ \tau}\begin{pmatrix}\|R^{t, t'}\| /\eta'\end{pmatrix}d\tau\\
=~&O(e^{\lambda^+ E}).
\end{split}
\end{equation}

\end{proof}

\jn{In the experimentation, we set the local epoch to a small number, e.g., 5, to reduce the deviation between local momentum and centralized optimization. }

\subsection{Adaptive Pruning}
\label{subsec:pruning}

In this section, we present our layer-adaptive pruning method, i.e., \ThePruneName{}. \ThePruneName{} considers the diverse features of each layer, the non-IID degrees of both the sensitive shared server data and the distributed sensitive device data, to remove useless filters in the convolutional layers while preserving the accuracy. \ThePruneName{} can reduce the communication overhead and the computational costs of the training process. \liu{Please note that \ThePruneName{} is performed only once on the server in a predefined round.} 

Algorithm \ref{alg:automatic_pruning} details \ThePruneName{}. On the server and each device (Lines \ref{line:localPRateBegin} - \ref{line:localPRateEnd}), we calculate a proper pruning rate based on the shared sensitive  server data or the distributed sensitive  device data (Line \ref{line:localPRate}). 
We denote the initial model $W_{k}$ on the server or Device $k$. At Round $T$, the updated model is denoted by $W'_{k}$. Then, the difference between the current model and the original one is $\Delta_{k} = W_{k} - W'_{k}$. Afterward, we can calculate the Hessian matrix, i.e., $H(W'_{k})$. We sort the eigenvalues of $H(W'_{k})$ in ascending order, i.e., $\{\lambda^m_{k}|m \in (1, d_{k})\}$ with $m$ referring the index of an eigenvalue and $d_{k}$ referring to the rank of the Hessian matrix. We take $B_{k}(\Delta_{k}) = H(W'_{k}) - \triangledown L(\Delta_{k} + W'_{k})$ as a base function with $\triangledown L(\cdot)$ corresponding to the gradients. We denote the Lipschitz constant of $B_{k}(\Delta_{k})$ as $\mathscr{L}_{k}$. Inspired by \cite{zhang2021validating}, we take the first $m_{k}$, which meets the condition of $\lambda_{m_{k+1}} - \lambda_{m_{k}} > 4\mathscr{L}_{k}$, to reduce accuracy degradation. Then, we can calculate the proper pruning rate by $p^*_{k} = \frac{m_{k}}{d_{k}}$, which is the ratio between the size of pruned eigenvalues and that of all the eigenvalues. As the proper pruning rate significantly differs in diverse devices due to the non-IID distribution, we utilize Formula \ref{eq:FedASP} to generate an aggregated proper pruning rate for the global model (Line \ref{line:fedPRateEnd}).
\begin{equation}
p^* = \sum_{k = 0}^{n} \frac{ \frac{n_k}{\mathcal{D}(P_k) + \epsilon}}{\sum_{k' = 0}^{n} \frac{n_{k'}}{\mathcal{D}(P_{k'}) + \epsilon}} * p^*_{k},
\label{eq:FedASP}
\end{equation}
where $\epsilon$ represents a small value to avoid the division of zero. 
We take a global threshold value ($\mathscr{V}$) calculated based on the aggregated proper pruning rate to serve as a baseline value for generating the pruning rate of each layer. $\mathscr{V}$ is the absolute value of the $\lfloor R*p^* \rfloor$-th smallest parameter in all the parameters (Lines \ref{line:globalSort} and \ref{line:threshold}).
Afterward, for each convolutional layer (Line \ref{line:cnnBegin}), we calculate the proper pruning rate by calculating the ratio between the number of parameters with inferior absolute values than $\mathscr{V}$ and the number of all the parameters (Lines \ref{line:localSort} - \ref{line:eachPRateEnd}).
Inspired by \cite{lin2020hrank}, we remove the filters corresponding to the smallest ranks in feature maps (the output of filters) based on the proper pruning rate in the layer (Lines \ref{line:rank} - \ref{line:replace}).
We take $R_l = \{r_l^j|j \in (1, d_l)\}$ to represent the ranks of feature maps at Layer $l$, where $d_l$ refers to the number of filters in this layer. 
As the feature maps are almost the same for a given model \cite{lin2020hrank}, we take the ranks calculated based on the server data and perform the pruning operations on the server because of its powerful computation capacity.
We calculate the feature maps in $R_l$ in Line \ref{line:rank} and sort them in Line \ref{line:sort}. 
We keep the filters having the last $d_l - \lfloor p^*_l * d_l \rfloor$ ranks in the sorted $R^l$, so as to attain the highest pruning rate $p_l \leq p^*_l$ (Line \ref{line:prune}). In the end, we take the preserved filters in the original model as the layer of the pruned model (Line \ref{line:replace}). When there is no available server data, the pruning process can also be performed on the server based on the rank values $R_l$ calculated using the dataset on a device, which are transferred to the server, with the execution of Line \ref{line:rank} being carried out at that device.

\begin{figure}[!t]
\centering
\subfigure[CNN]{
\includegraphics[width=0.23\linewidth]{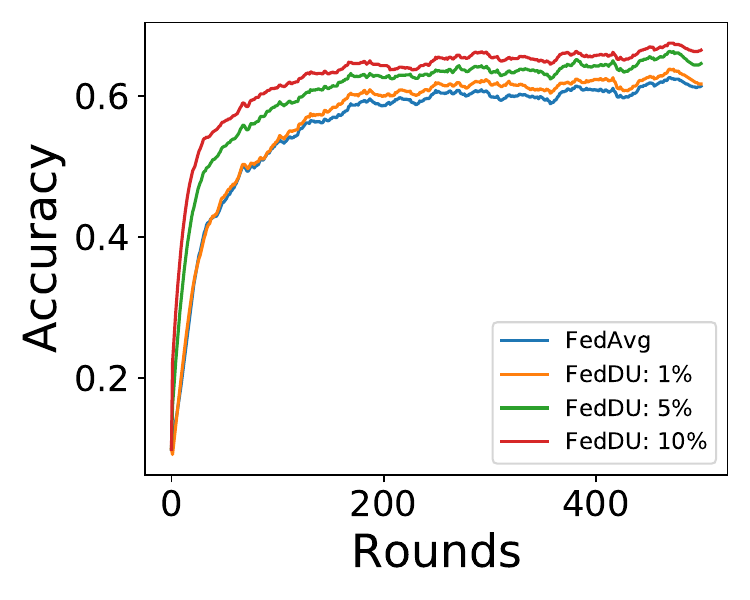}
}
\subfigure[VGG]{
\includegraphics[width=0.23\linewidth]{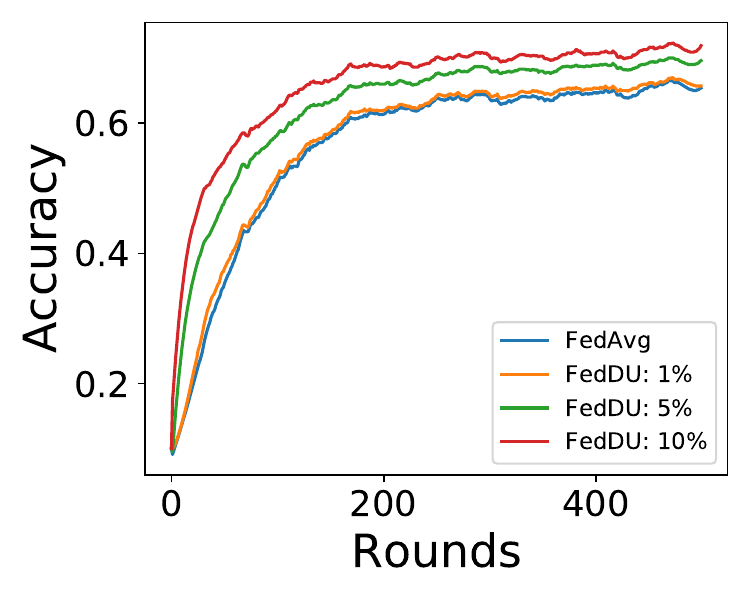}
}
\caption{\textbf{The accuracy of \TheAlgoName{} with diverse amounts of server data on CIFAR-10.} $1\%$, $5\%$, $10\%$ represent the value of $p$ (see details in Section 4.1
).}
\label{fig:dynamic}
\end{figure}

\begin{figure}[!t]
\centering
\subfigure[CNN]{
\includegraphics[width=0.23\linewidth]{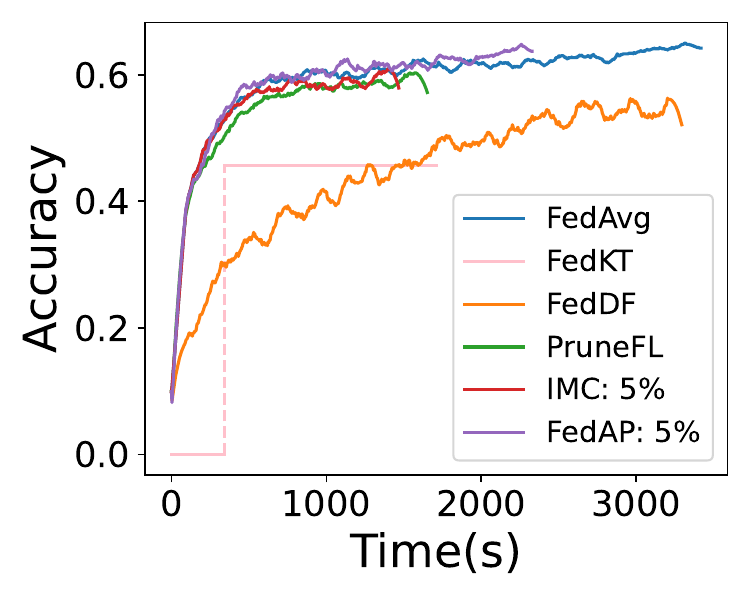}
}
\subfigure[CNN]{
\includegraphics[width=0.23\linewidth]{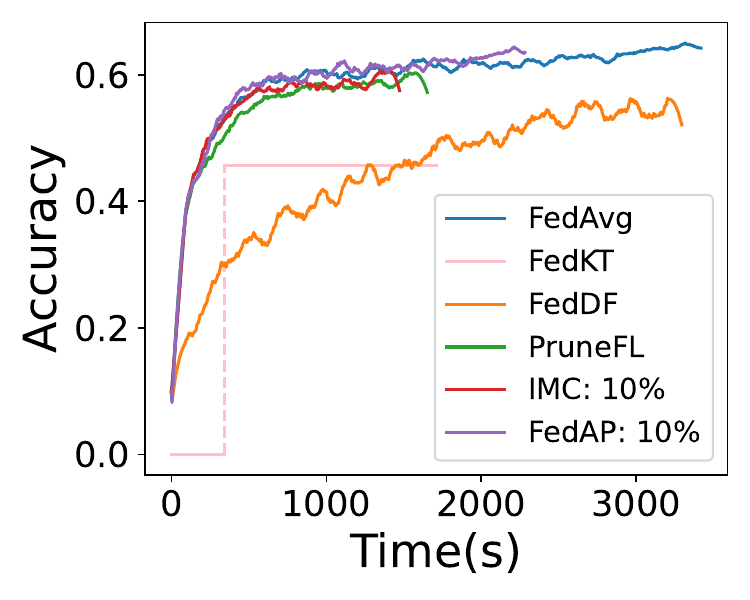}
}
\subfigure[ResNet]{
\includegraphics[width=0.23\linewidth]{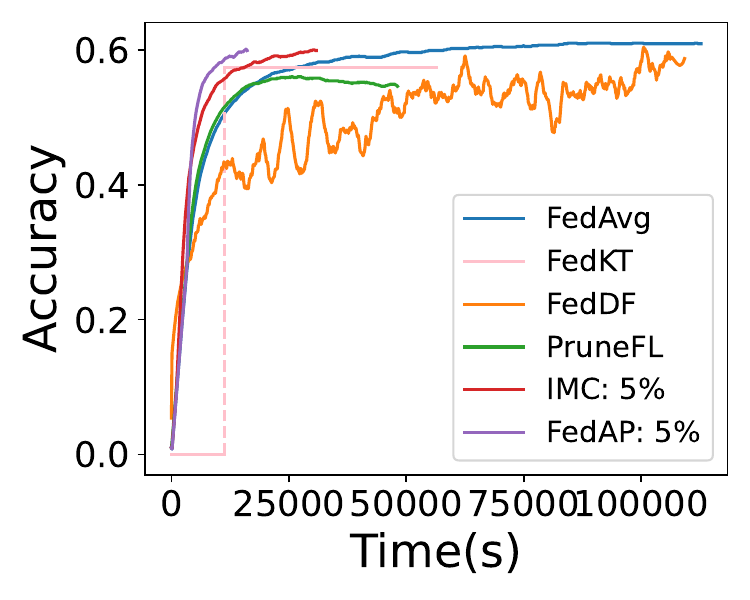}
}
\subfigure[ResNet]{
\includegraphics[width=0.23\linewidth]{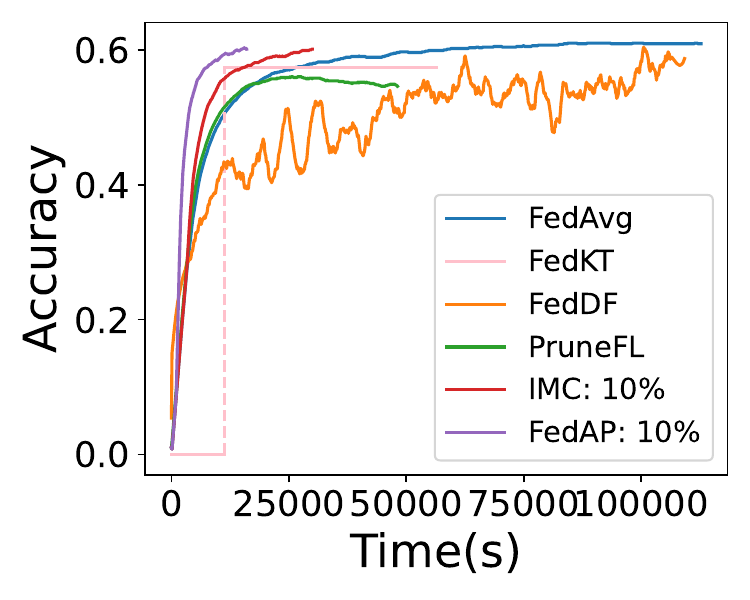}
}
\caption{\textbf{The accuracy and training time with diverse model update methods corresponding to \TheAlgoName{} with $p = 5\%$ and $p = 10\%$.} CNN is with CIFAR-10 and ResNet is with CIFAR-100.}
\label{fig:cmp_share_l5}
\end{figure}

\section{Experiments}
\label{sec:experiments}

In this section, we demonstrate the experimental results to show the advantages of \TheNameM{} by comparing it with state-of-the-art baselines, i.e., FedAvg~\cite{mcmahan2017communication}, \rev{FedKT \cite{li2021practical}, FedDF \cite{lin2020ensemble}}, Data-sharing \cite{zhao2018federated},  Hybrid-FL \cite{yoshida2020hybrid}, \jn{server-side momentum \cite{reddi2018adaptive}, device-side momentum \cite{karimireddy2020mime}, FedDA \cite{jin2022accelerated},} HRank \cite{lin2020hrank}, IMC~\cite{zhang2021validating}, and PruneFL \cite{jiang2023model}.

\subsection{Experimental Setup}
\label{subsec:expSetup}

We evaluate \TheNameM{} using an FL system consisting of a parameter server and $100$ devices. In each round, we randomly choose $10$ devices. We utilize two real-life datasets, i.e., CIFAR-10 and CIFAR-100 \cite{krizhevsky2009learning}, in the experimentation. We exploit four models, i.e., a simple synthetic CNN network (CNN), ResNet18 (ResNet) \cite{he2016deep}, VGG11 (VGG) \cite{simonyan2014very}, and LeNet5 (LeNet) \cite{lecun1998gradient}. 
The experimentation of CNN, VGG, and LeNet is carried out on both CIFAR-10 and CIFAR-100, while that of ResNet is performed on CIFAR-100. 

CNN consists of three $3*3$ convolution layers, a fully connected layer, and a final softmax output layer. The first layer contains 32 channels, while the second and third layers have 64 channels.
The first and the second layers are followed by $2 * 2$ max pooling. The fully connected layer contains $64$ units and exploits ReLu activation. CNN contains 122570 parameters in total. The mini-batch size is set to $10$ for the local model update, and the selected device executes $E = 5$ local epochs. The local learning rate $\eta$ is $0.1$ with a decay rate of $0.99$. $\mathcal{C}$ is set to 1. The experimentation is carried out using 33 Tesla V100 GPUs to simulate an FL environment composed of a parameter server and 100 devices. 
\liu{We carry out the pruning process in Round 30 in all the experimentation with adaptive pruning.}
In addition, we exploit the Savitzky–Golay filter \cite{press1990savitzky} to smooth the accuracy in figures.

The CIFAR-10 dataset contains 60000 images (50000 for training and 10000 for testing), each corresponding to one of ten categories. We take 40000 images in the training dataset as device data and randomly select $p$ * 40000 images from the remaining 10000 images as server data, with $0\% < p < 25\%$. $p$ represents the ratio between the size of the shared insensitive server and that of the distributed sensitive device data. For the non-IID setting, we sort the device data according to the label and then divide these data evenly into $2 * 100$ fractions. Each device is randomly assigned $2$ fractions, and most devices contain the data with 2 labels. We utilize the same method to handle the CIFAR-100 dataset.


\begin{table}[!t]
  \centering
  \small
    \caption{\textbf{The accuracy with diverse effective steps on CIFAR-10.} ``$\mathcal{A}$'' represents FedAvg. ``$\mathcal{D}$'' represents \TheAlgoName{}. ``$\mathcal{S}$'' represents \TheAlgoName{}-S.
    }
    \begin{tabular}{|c|c|c|c|c|c|c|c|c|c|c|}
    \hline
    \multirow{2}[4]{*}{Method} & \multirow{2}[4]{*}{$p$} & \multicolumn{4}{c|}{$\tau_{eff}$ (CNN)} &  
    \multicolumn{4}{c|}{$\tau_{eff}$ (VGG)} \bigstrut\\
\cline{3-10}     &   & 5    & 10    & 20    & $\frac{n_0E}{B}$ & 5    & 10    & 20    & $\frac{n_0E}{B}$ \bigstrut\\
    \hline
    $\mathcal{A}$ & & \multicolumn{4}{c|}{0.626}  & \multicolumn{4}{c|}{0.666} \bigstrut\\
\cline{1-10} $\mathcal{D}$ & \multirow{0}[2]{*}{1\%} & \multicolumn{4}{c|}{\textbf{0.638}} & \multicolumn{4}{c|}{\textbf{0.670}} \bigstrut\\
\cline{1-1}\cline{3-10} $\mathcal{S}$ & & 0.628 & \textbf{0.638} & 0.631 & 0.577 & 0.668 & \textbf{0.672} & 0.668 & 0.627 \bigstrut\\
\cline{1-10} $\mathcal{D}$ & \multirow{0}[2]{*}{5\%} & \multicolumn{4}{c|}{\textbf{0.663}} & \multicolumn{4}{c|}{\textbf{0.700}} \bigstrut\\
\cline{1-1}\cline{3-10} $\mathcal{S}$ & & 0.636 & 0.639 & \textbf{0.648} & 0.436 & 0.674 & 0.684 & \textbf{0.690}  & 0.651 \bigstrut\\
\cline{1-10}  $\mathcal{D}$ & \multirow{0}[2]{*}{10\%} & \multicolumn{4}{c|}{\textbf{0.675}} & \multicolumn{4}{c|}{\textbf{0.723}} \bigstrut\\
\cline{1-1}\cline{3-10} $\mathcal{S}$ & & 0.640 & 0.643 & \textbf{0.658} & 0.429 & 0.682 & 0.689 & \textbf{0.707} & 0.664 \bigstrut\\
    \hline
    \end{tabular}%
  \label{tab:basic}%
\end{table}%

\begin{figure}[!t]
\centering
\subfigure[VGG]{
\includegraphics[width=0.23\linewidth]{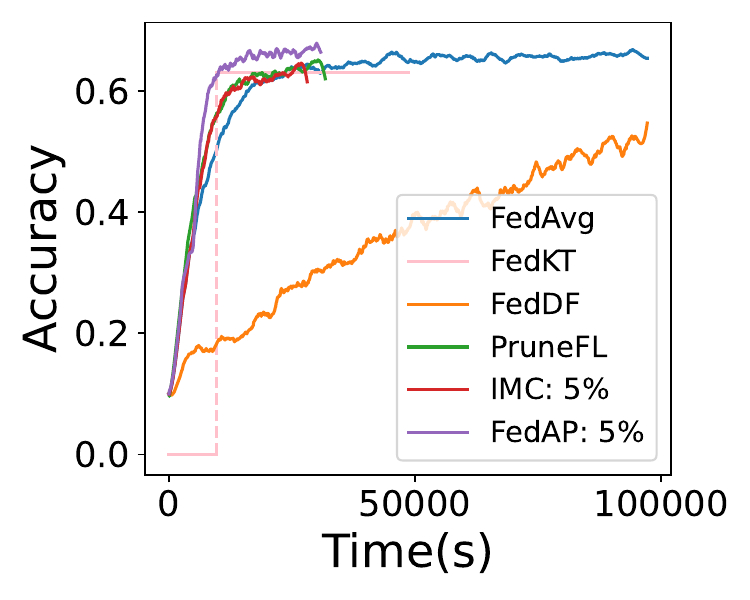}
}
\subfigure[VGG]{
\includegraphics[width=0.23\linewidth]{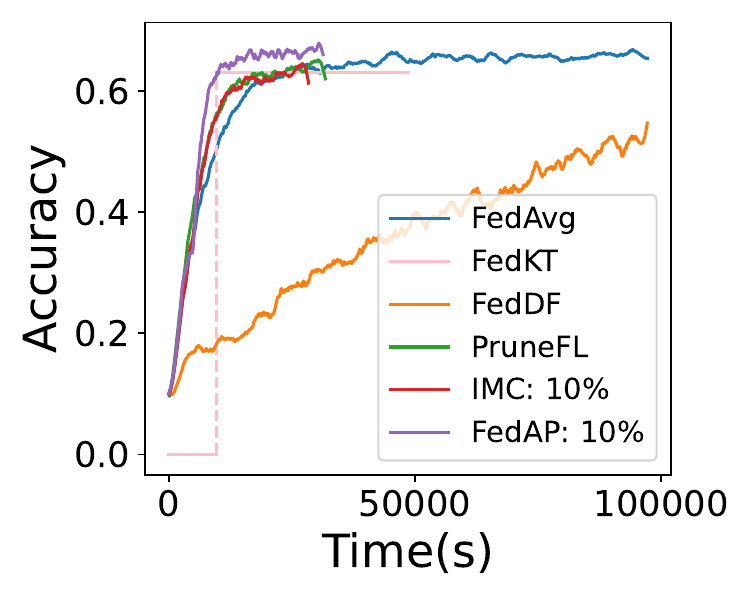}
}
\subfigure[LeNet]{
\includegraphics[width=0.23\linewidth]{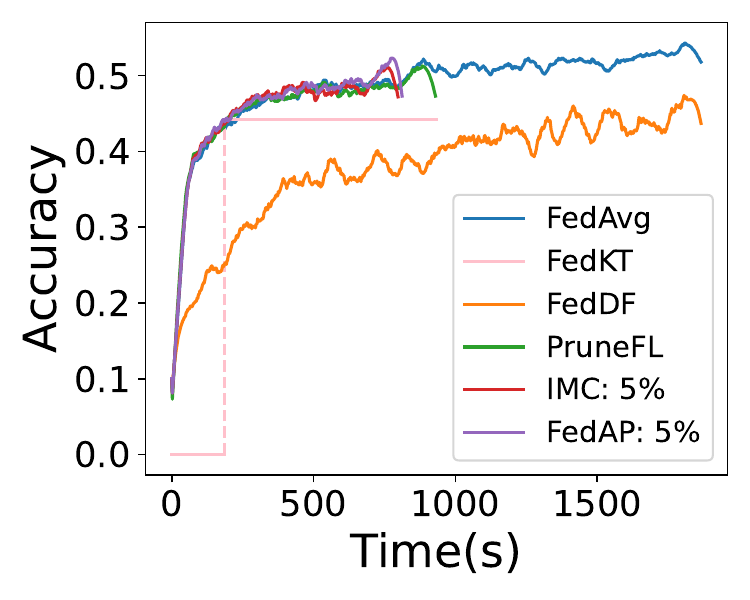}
}
\subfigure[LeNet]{
\includegraphics[width=0.23\linewidth]{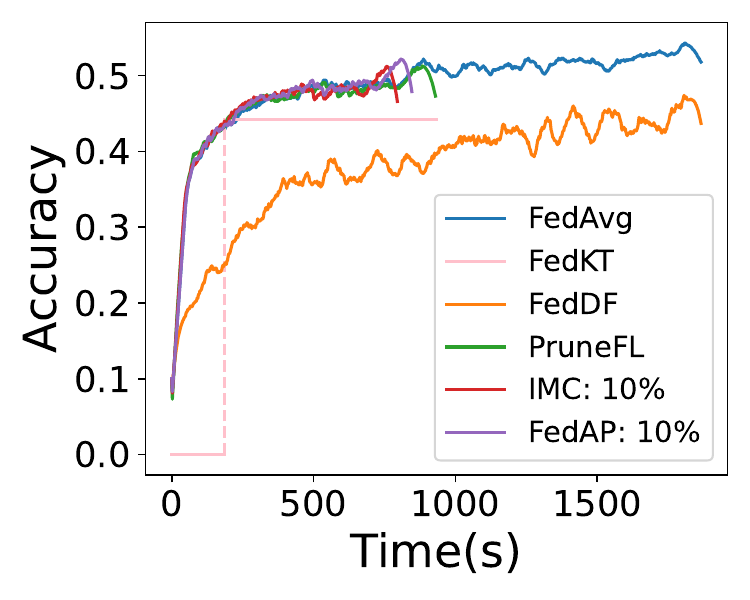}
}
\caption{\textbf{The accuracy and training time with diverse model update methods corresponding to \TheAlgoName{} based on CIFAR-10.}}
\label{fig:cmp_share_l5_cifar10}
\end{figure}

\subsection{Evaluation with non-IID Data}

We first conduct comparison of \TheAlgoName{} with FedAvg, \rev{FedKT, FedDF,} Data-sharing, and Hybrid-FL in terms of model accuracy. \jn{Then, we compare \TheAlgoNameM{} with server-side momentum, device-side momentum, and FedDA.} Afterward, we show the advantages of the adaptive pruning method, i.e., \ThePruneName{}, with the comparison of FedAvg, IMC, HRank, and PruneFL, in terms of model efficiency. \jn{Last but not least, we demonstrate that \TheNameM{}, which 
consists of both \TheAlgoNameM{} and \ThePruneName{}, significantly outperforms the eight state-of-the-art baselines in terms of accuracy, efficiency and computational cost. Finally, we present our ablation study.}

\begin{table}[!t]
  \centering
  {\small
  \caption{\textbf{The accuracy with $f'(acc) = 1 - acc$ and $f'(acc) = \frac{1}{acc + \epsilon}$ on CIFAR-10.}}
    \begin{tabular}{|c|c|c|c|c|c|c|}
    \hline
    \multirow{3}[4]{*}{Method} & \multirow{3}[4]{*}{$p$} & \multicolumn{2}{c|}{CNN} & \multicolumn{2}{c|}{VGG} \bigstrut\\
\cline{3-6}
    & & \multicolumn{2}{c|}{Accuracy} 
& \multicolumn{2}{c|}{Accuracy} \bigstrut\\
\cline{3-6}         &       & $1-acc$ & $\frac{1}{acc + \epsilon}$ & $1-acc$ & $\frac{1}{acc + \epsilon}$ \bigstrut\\
    \hline
    FedAvg &       & \multicolumn{2}{c|}{0.626} & \multicolumn{2}{c|}{0.666}  \bigstrut\\
\cline{1-6}  \multirow{3}[6]{*}{\TheAlgoName{}} & 1\%   & \textbf{0.638} & 0.637 & 0.670  & \textbf{0.677} \bigstrut\\
\cline{2-6}        & 5\%   & \textbf{0.663} & 0.660 & \textbf{0.700} & 0.697 \bigstrut\\
\cline{2-6}       & 10\%  & \textbf{0.675} & 0.668 & \textbf{0.723} & 0.711 \bigstrut\\
    \hline
    \end{tabular}%
  \label{tab:cmp_minus_divide}%
  }
\end{table}%

\begin{figure}[!t]
\centering
\subfigure[CNN]{
\includegraphics[width=0.23\linewidth]{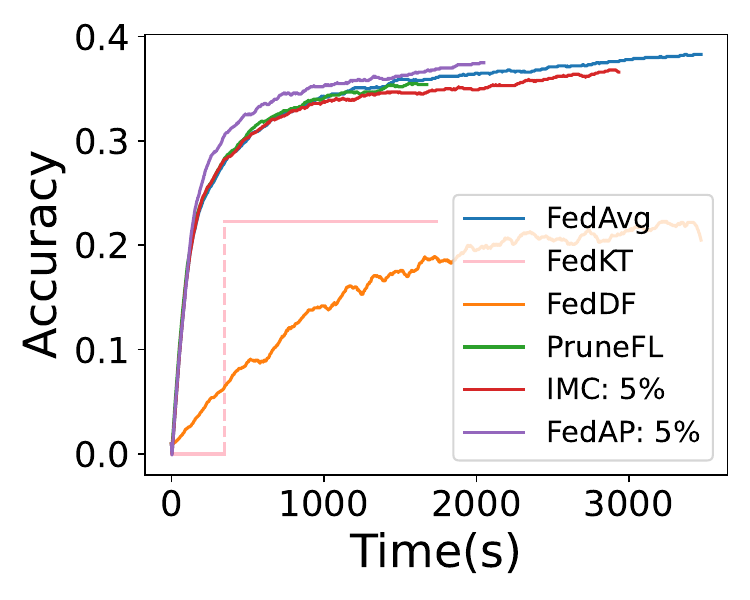}
}
\subfigure[CNN]{
\includegraphics[width=0.23\linewidth]{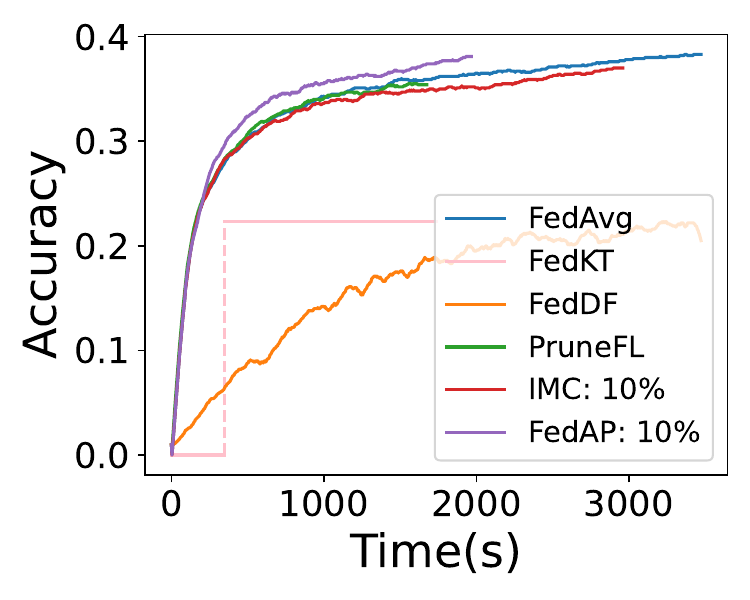}
}
\subfigure[VGG]{
\includegraphics[width=0.23\linewidth]{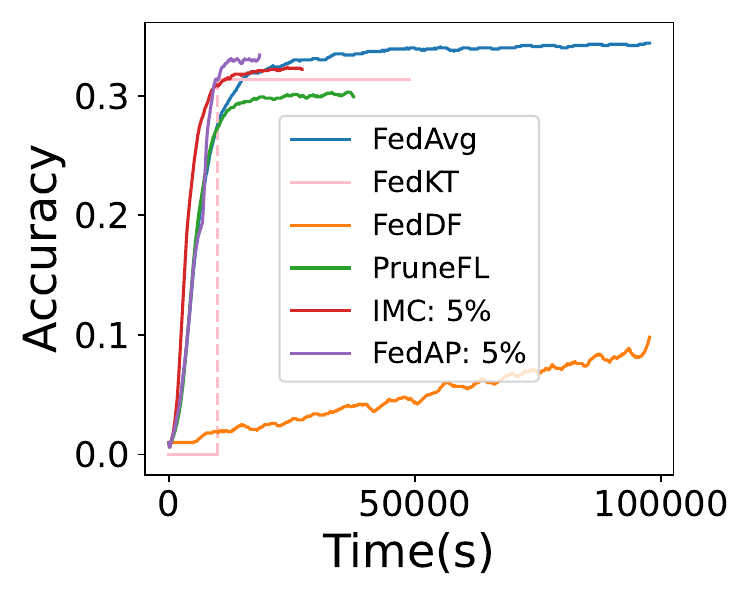}
}\\
\subfigure[VGG]{
\includegraphics[width=0.23\linewidth]{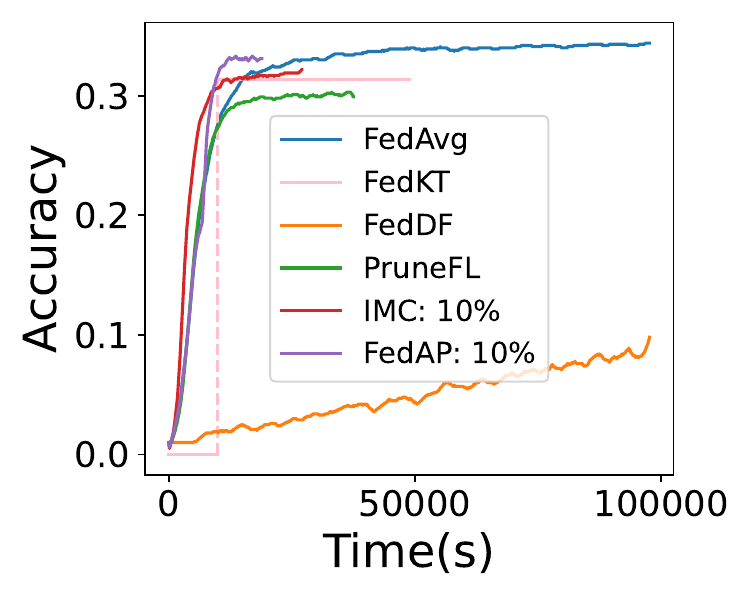}
}
\subfigure[LeNet]{
\includegraphics[width=0.23\linewidth]{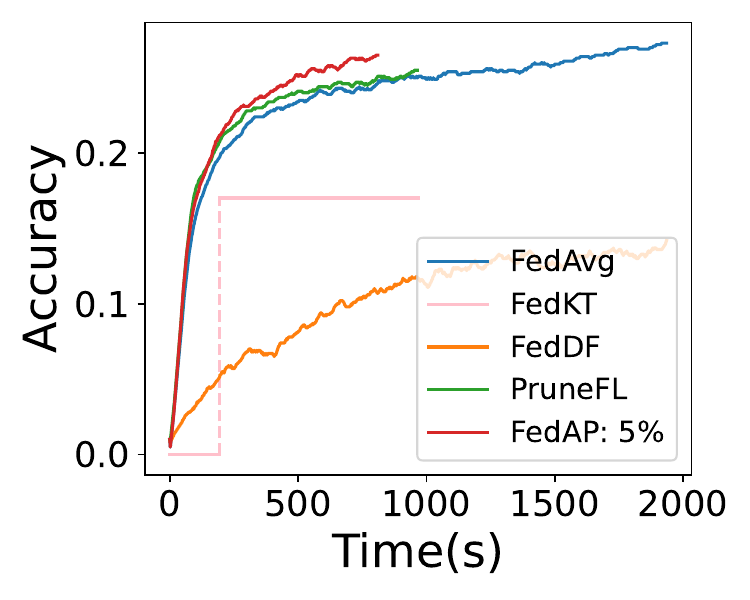}
}
\subfigure[LeNet]{
\includegraphics[width=0.23\linewidth]{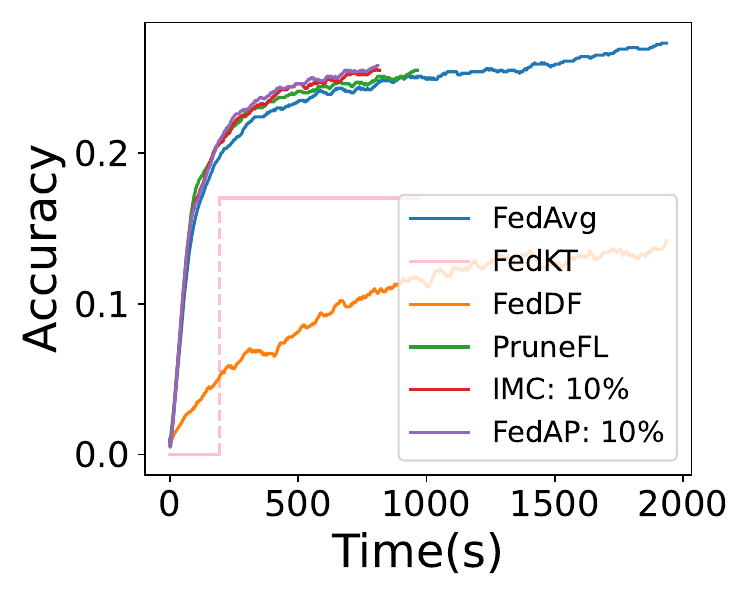}
}
\caption{\textbf{The accuracy and training time with diverse model update methods corresponding to \TheAlgoName{} based on CIFAR-100.}}
\label{fig:cmp_share_l5_cifar100}
\end{figure}

\subsubsection{Evaluation on \TheAlgoName{}}

\jn{In this part, we compare \TheAlgoName{} with five baseline methods, i.e., FedAvg, \rev{FedKT, FedDF,} Data-sharing, and Hybrid-FL. Afterward, we analyze the effect of $\tau_{eff}$, the effect of $f'(acc)$, the effect of $\mathcal{C}$, and the effect of the non-IID degree of server data.}
Data-sharing transfers the shared server data to devices in order to combine the local data and the server data, which may have a smaller non-IID degree compared with the original local data, so as to improve the accuracy of the updated model on the device. Hybrid-FL takes the shared server data of significant size as an ordinary dataset on a device and utilizes the FedAvg algorithm to perform the training process.

\jn{In order to take advantage of the server data, we have significant effective steps at the beginning, which leads to a quick increase of accuracy. Then, at the end of the training, we reduce the effective steps and focus on the device data to achieve high accuracy. }
Figure \ref{fig:dynamic} shows that \TheAlgoName{} corresponds to excellent efficiency and high accuracy compared with baseline methods with CIFAR-10. In addition, with more server data, \TheAlgoName{} can achieve better performance in terms of accuracy (up to 5.3\% higher) for both CNN and VGG. As shown in Figures \ref{fig:cmp_share_l5} and \ref{fig:cmp_share_l5_cifar10}, \TheAlgoName{} leads to a higher accuracy compared with FedAvg (up to 5.7\%
), \rev{FedKT (up to 22.6\%), FedDF (up to 5.0\%)}, Data-sharing (up to 11.7\%
), and Hybrid-FL (up to 19.5\%
) for both CNN and ResNet when $p = 5\%$ and $10\%$. In contrast, Data-sharing has slightly higher accuracy (up to 1.2\%) compared with \TheAlgoName{} for VGG, as the more complex model can be better trained with augmented data. 
\jn{Compared with \TheAlgoName{}, Data-sharing needs to transfer the server data to devices, which may incur privacy issues and corresponds to high communication overhead. In addition, Data-sharing leads to a much longer training time to achieve the accuracy of 0.6 (for CNN, up to 15.7 times slower) and 0.4 (for LeNet, up to 28.9 times slower) than \TheAlgoName{} on CNN.}

\begin{table}[!t]
  \centering
  {\small
  \caption{\textbf{The accuracy with diverse values of $\mathcal{C}$ on CIFAR-10.}}
    \begin{tabular}{|c|c|c|c|c|c|c|c|c|}
    \hline
    \multirow{2}[4]{*}{Method} & \multirow{2}[4]{*}{$p$} & \multicolumn{3}{c|}{$\mathcal{C}$ (CNN)} & \multicolumn{3}{c|}{$\mathcal{C}$ (VGG)} \bigstrut\\
\cline{3-8}       &       &     1.5   & 1     & 0.5 &     1.5   & 1     & 0.5\bigstrut\\
    \hline
    FedAvg &       & \multicolumn{3}{c|}{0.626} & \multicolumn{3}{c|}{0.666} \bigstrut\\
\cline{1-8} \multirow{3}[6]{*}{\TheAlgoName{}} & 1\%   & 0.636 & \textbf{0.638} & 0.634 & \textbf{0.673} & 0.670  & 0.672 \bigstrut\\
\cline{2-8}       & 5\%   & 0.651 & \textbf{0.663} & 0.647 & 0.685 & \textbf{0.700} & 0.690 \bigstrut\\
\cline{2-8}       & 10\%  & 0.669 & \textbf{0.675} & 0.668 & \textbf{0.724} & 0.723 & 0.715 \bigstrut\\
    \hline
    \end{tabular}%
  \label{tab:diverse_c}%
  }
\end{table}%

\begin{figure}[!t]
\centering
\subfigure[CNN]{
\includegraphics[width=0.23\linewidth]{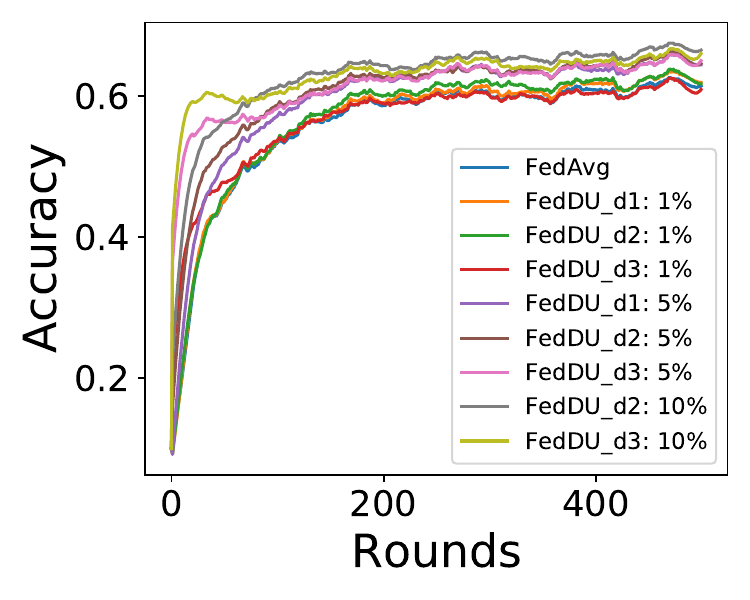}
}
\subfigure[VGG]{
\includegraphics[width=0.23\linewidth]{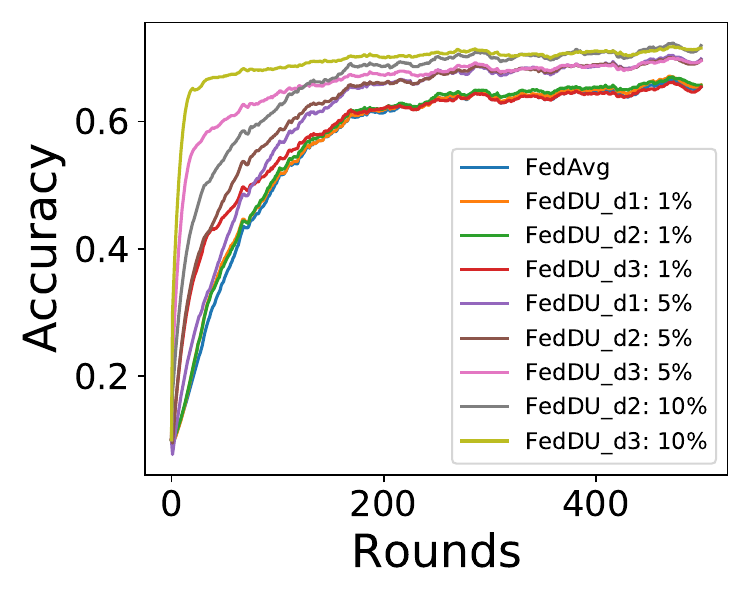}
}
\caption{\textbf{The accuracy of \TheAlgoName{} with the server data of diverse non-IID degrees on CIFAR-10.} $1\%$, $5\%$, $10\%$ represent the value of $p$ (see details in Section 4.1). 
``$d$'' represents the non-IID degree. $d1 = 0.61$, $d2 = 0.31$ and $d3 = 9.0 * 10^{-6}$.}
\label{fig:cmp_l5_l10}
\end{figure}

\begin{table}[!t]
  \centering
  {\small
    \caption{\textbf{The accuracy of \TheAlgoName{} with the server data of diverse non-IID degrees on CIFAR-10.} ``$d$'' represents the non-IID degree . See details of $p$ in Section 4.1.
    }
    \begin{tabular}{|c|c|c|c|c|c|c|c|c|}
    \hline
    \multirow{2}[4]{*}{Method} & \multirow{2}[4]{*}{$p$} & \multicolumn{3}{c|}{$d$ (CNN)} & \multicolumn{3}{c|}{$d$ (VGG)} \bigstrut\\
\cline{3-8}        &       & $0.61$  & $0.31$  & $9.0 * 10^{-6}$& $0.61$  & $0.31$  & $9.0 * 10^{-6}$ \bigstrut\\
    \hline
    FedAvg &       & \multicolumn{3}{c|}{0.626} & \multicolumn{3}{c|}{0.666} \bigstrut\\
\cline{1-8} \multirow{3}[6]{*}{\TheAlgoName{}} & 1\%   & 0.635 & \textbf{0.638} & 0.625 & \textbf{0.671} & 0.670  & 0.662 \bigstrut\\
\cline{2-8}       & 5\%   & 0.659 & \textbf{0.663} & 0.659 & \textbf{0.704} & 0.700   & 0.700 \bigstrut\\
\cline{2-8}       & 10\%  &   0.645    & \textbf{0.675} & 0.667 & 0.696    & \textbf{0.723} & 0.718 \bigstrut\\
    \hline
    \end{tabular}%
  \label{tab:cmp_l5_l10}%
    }
\end{table}%

\jn{We have similar results on CIFAR-100 as that of CIFAR-10. As shown in Figure \ref{fig:cmp_share_l5_cifar100}, \TheAlgoName{} corresponds to a higher accuracy compared with Data-sharing (up to 14.0\%), \rev{FedKT (up to 16.0\%), FedDF (up to 17.0\%)}, Hybrid-FL (up to 20.4\%) and FedAvg (up to 2.0\%) with CNN, VGG, LeNet and ResNet, when $p = 5\%$ and $p = 10\%$.  In addition, Data-sharing suffers from a much longer training time to achieve the accuracy of 0.2 (up to 25.2 times slower) than \TheAlgoName{}.}


\jn{\textbf{Effect of \texorpdfstring{$\tau_{eff}$}{teff}}: 
First, we fix the number of effective steps, and we denote this setting \TheAlgoName{}-Static (\TheAlgoName{}-S). We carry out the experiments with diverse values, i.e., $\tau_{eff} = 5$, $\tau_{eff} = 10$, $\tau_{eff} = 20$ and $\tau_{eff} = \frac{n_0E}{B} > 20$. As shown in Table \ref{tab:basic}, when the effective number of steps is small, i.e., $\tau_{eff} = 5$ or $10$, the corresponding accuracy is higher than FedAvg with a small margin, e.g., up to 1.7\% for CNN and 2.3\% for VGG.  
However, when the number of effective steps is huge, e.g., $\frac{n_0E}{B} > 200$, the accuracy becomes the lowest. This is expected as the global model is updated too much towards the central data, which degrades the final performance. We find that dynamical adjustment of $\tau_{eff}$, i.e., \TheAlgoName{}, can significantly improve the accuracy (up to 1.7\% higher accuracy compared with the static method).}

\begin{table}[!t]
  \centering
  {\small
  \caption{\textbf{The final accuracy, training time, and computational cost of VGG with FedAvg, IMC, PruneFL, and \ThePruneName{} based on CIFAR-10.}}
    \begin{tabular}{|c|c|c|c|c|}
    \hline
    Methods & Accuracy & Time(0.6) &  MFLOPs \bigstrut\\
    \hline
    FedAvg & 0.666 & 15953 & 153.4 \bigstrut\\
\cline{1-4} IMC   & 0.645 & 11992 & 153.4 \bigstrut\\
\cline{1-4} PruneFL & 0.652 & 11896 & 153.4 \bigstrut\\
\cline{1-4} \ThePruneName{} & \textbf{0.670} & \textbf{7980} & \textbf{56.1} \bigstrut\\
    \hline
    \end{tabular}%
  \label{tab:vgg_prune_cmp_cifar10}%
  }
\end{table}%

\begin{figure}[!t]
\centering
\subfigure[CNN]{
\includegraphics[width=0.23\linewidth]{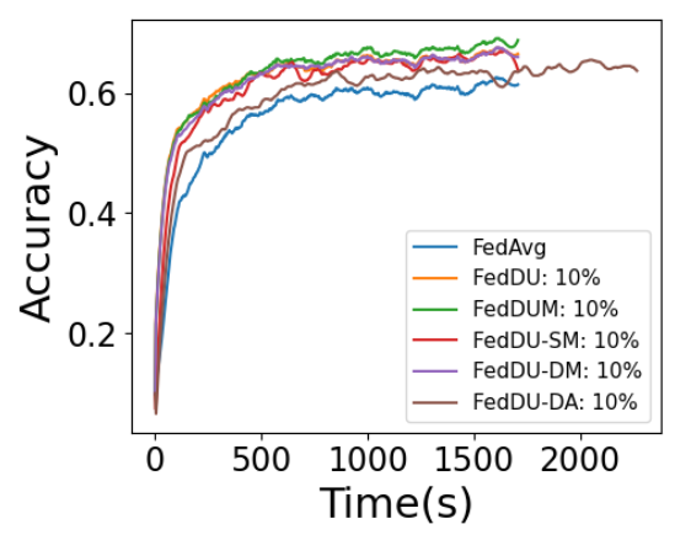}
}
\subfigure[VGG]{
\includegraphics[width=0.23\linewidth]{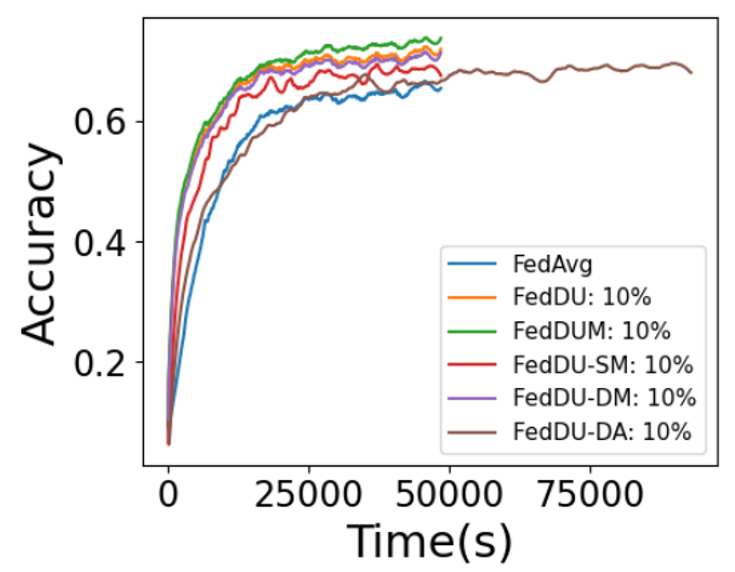}
}
\subfigure[LeNet]{
\includegraphics[width=0.23\linewidth]{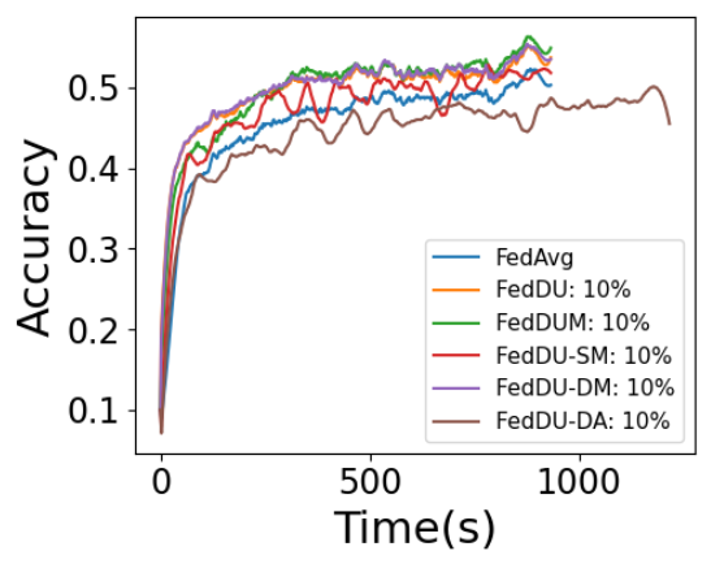}
}
\caption{\textbf{The accuracy of \TheAlgoNameM{} with diverse adaptive optimization methods on CIFAR-10.} }
\label{fig:fedDUM_10}
\end{figure}

\jn{\textbf{Effect of \texorpdfstring{$f'(acc)$}{f(acc)}}:
We can choose a synthetic function of $f'(acc)$ between $1 - acc$ and $\frac{1}{acc + \epsilon}$. As shown in Table \ref{tab:cmp_minus_divide}, we empirically find that the performance corresponding to $1 - acc$ is similar to that corresponding to $\frac{1}{acc + \epsilon}$ in terms of accuracy, while $1 - acc$ is slightly better (0.32\% on average). Thus, we choose $1 - acc$ in \TheAlgoName{}.}

\jn{\textbf{Effect of \texorpdfstring{$\mathcal{C}$}{C}}: In order to choose an appropriate value of $\mathcal{C}$, we empirically analyze the performance corresponding to diverse values of $\mathcal{C}$, i.e., 1.5, 1, 0.5. As shown in Table \ref{tab:diverse_c}, we find that the performance of $\mathcal{C}= 1$ has the highest accuracy on average. Then, we choose $\mathcal{C}=1$ in \TheAlgoName{}.}

\jn{\textbf{Influence of the Non-IID Degree of Server Data}: The non-IID degree of server data can have a significant influence on the training process. When the non-IID degree is small, the distribution of the server data is similar to that of the global distribution of all the device data, which is beneficial to the training process. Otherwise, the server data is of less help. In order to analyze the influence of the non-IID degree of the server data, we carry out the experiment with diverse non-IID degrees, e.g., $0.61$, $0.31$ and $9.0*10^{-6}$. As shown in Figure \ref{fig:cmp_l5_l10}, when the non-IID degree is smaller, the training is more efficient and it takes much shorter time to achieve a target accuracy (up to 6.6 times faster to achieve the accuracy of 0.6). However, as shown in Table \ref{tab:cmp_l5_l10}, the accuracy is not directly related to the non-IID degrees.}

\begin{table}[!t]
  \centering
  \caption{\textbf{The final accuracy, training time, and computational cost of LeNet with FedAvg, IMC, PruneFL, and \ThePruneName{} based on CIFAR-10.}}
    \begin{tabular}{|c|c|c|c|c|}
    \hline
     Methods & Accuracy & Time(0.49) &  MFLOPs \bigstrut\\
    \hline
     FedAvg & \textbf{0.523} & 502   & 0.7 \bigstrut\\
\cline{1-4}          IMC   & 0.513 & 583   & 0.7 \bigstrut\\
\cline{1-4}          PruneFL & 0.515 & 529   & 0.7 \bigstrut\\
\cline{1-4}          \ThePruneName{} & \textbf{0.522} & \textbf{483} & \textbf{0.6} \bigstrut\\
    \hline
    \end{tabular}%
  \label{tab:lenet_prune_cmp_cifar10}%
\end{table}%

\begin{figure}[!t]
\centering
\subfigure[CNN]{
\includegraphics[width=0.23\linewidth]{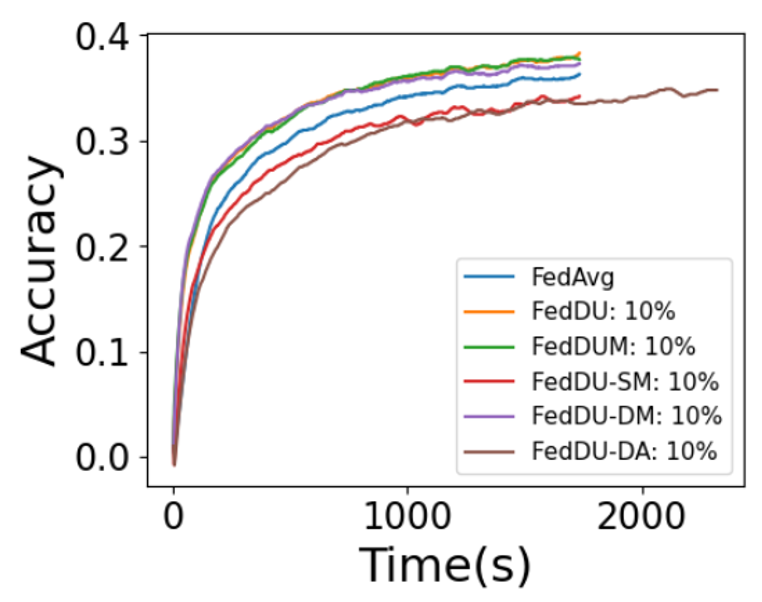}
}
\subfigure[VGG]{
\includegraphics[width=0.23\linewidth]{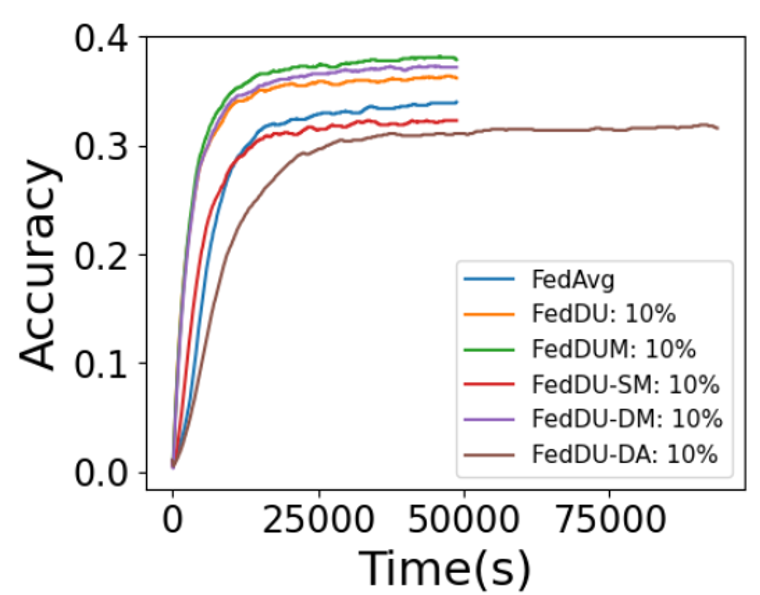}
}
\subfigure[LeNet]{
\includegraphics[width=0.23\linewidth]{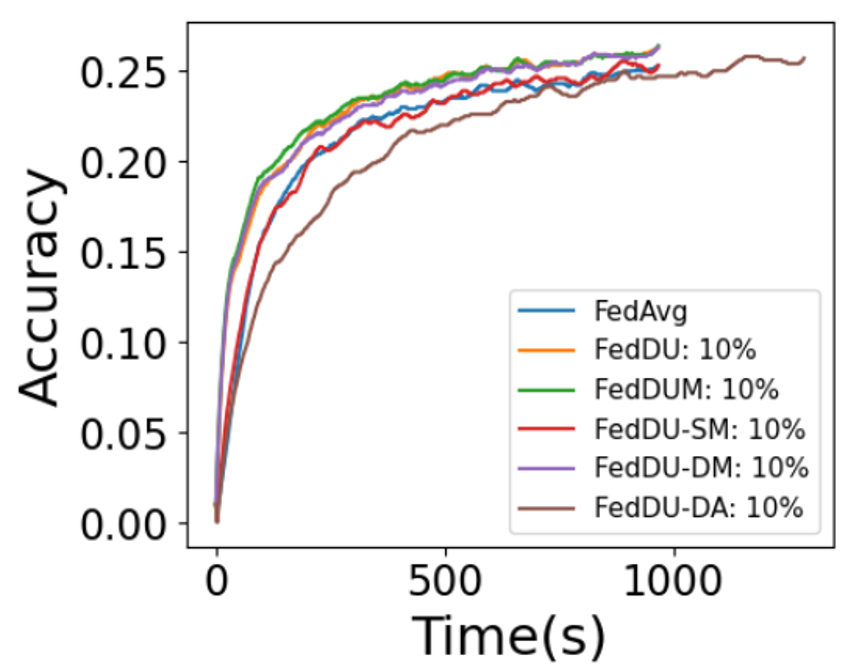}
}
\caption{\textbf{The accuracy of \TheAlgoNameM{} with diverse adaptive optimization methods on CIFAR-100.} }
\label{fig:fedDUM_100}
\end{figure}

\subsubsection{\jn{Evaluation on \TheAlgoNameM{}}}

\jn{In this section, we compare \TheAlgoNameM{} with two adapted adaptive optimization methods, i.e., server-side momentum (FedDU-SM) \cite{reddi2018adaptive} and device-side momentum (FedDU-DM) \cite{karimireddy2020mime}, as well as FedDA (FedDU-DA) \cite{jin2022accelerated}. For a fair comparison, each optimization method is combined with FedDU to utilize the server data in order to improve the accuracy. }

\jn{As shown in Figure \ref{fig:fedDUM_10}, compared with FedAvg and \TheAlgoName{}, \TheAlgoNameM{} achieves significantly higher accuracy within the same training time for CNN (up to 6.5\% for FedAvg, 1.6\% for \TheAlgoName{}), VGG (up to 7.2\% for FedAvg, 1.5\% for \TheAlgoName{}), and LeNet (up to 4.0\% for FedAvg, 1.1\% for \TheAlgoName{}) on CIFAR-10. \TheAlgoNameM{} leads to a higher accuracy compared with FedDU-SM (up to 4.4\%), FedDU-DM (up to 1.6\%), and FedDU-DA (up to 6.2\%). In addition, FedDU-DA corresponds to a longer total training time (up to 91.8\% compared with other methods) because of additional communication cost. Furthermore, FedDUM corresponds to a shorter training time to achieve the target accuracy (up to 1.9 times faster than FedAvg, 44.2\% faster than FedDU-SM, 42.9\% faster than FedDU-DM and 3.1 times faster than FedDA). }

\begin{figure}[t]
\centering
\subfigure[CNN]{
\includegraphics[width=0.3\linewidth]{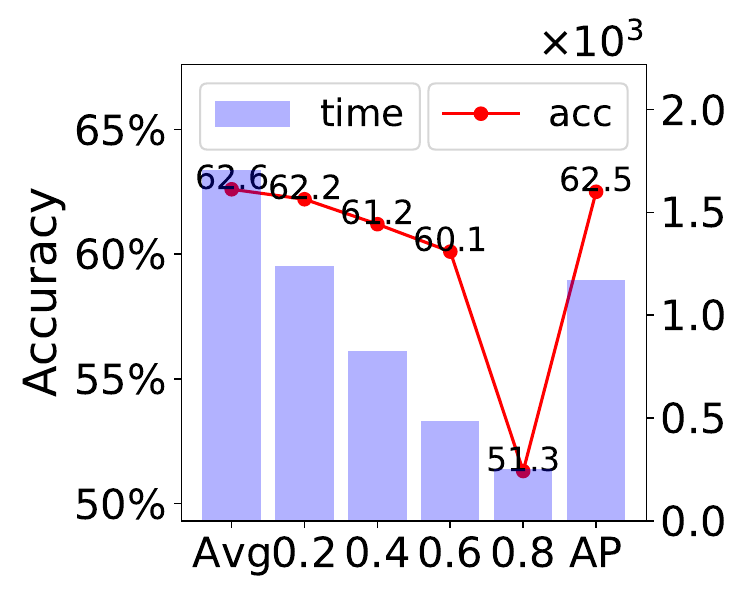}
}
\subfigure[VGG]{
\includegraphics[width=0.3\linewidth]{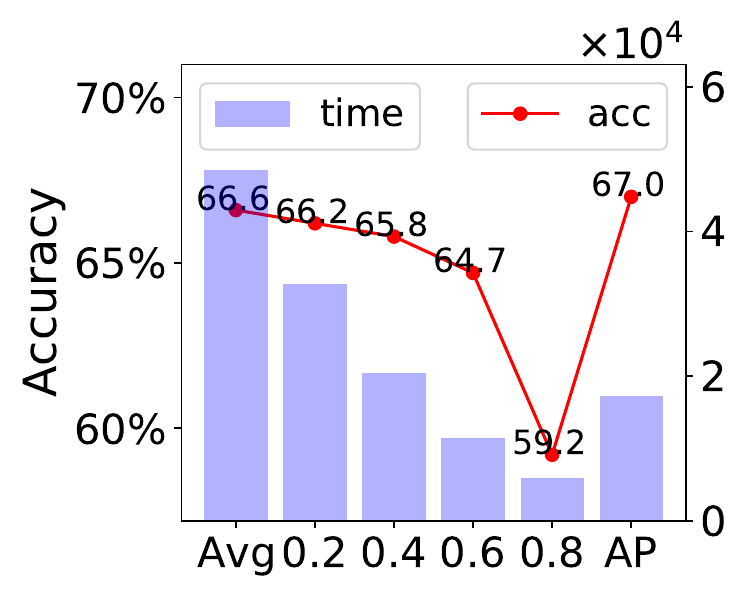}
}
\caption{\textbf{The accuracy and the training time with FedAvg, HRank of diverse pruning rates, and \ThePruneName{} on CIFAR-10.}}
\label{fig:cmp_prunerate_l5}
\end{figure}

\begin{table}[!t]
  \centering
  \small
  \caption{\textbf{The final accuracy, training time, and computational cost with FedAvg, IMC, PruneFL, and \ThePruneName{} based on CIFAR-100.}}
    \begin{tabular}{|c|c|c|c|c|c|c|}
    \hline
    \multirow{2}[4]{*}{Method} & \multicolumn{3}{c|}{CNN} & \multicolumn{3}{c|}{VGG} \bigstrut\\
\cline{2-7}     & Accuracy & Time(0.3) &  MFLOPs & Accuracy & Time(0.3) &  MFLOPs \bigstrut\\
    \hline
    FedAvg & \textbf{0.363} & 494   & 4.6 & \textbf{0.340} & 12798 & 153.5 \bigstrut\\
\cline{1-7} IMC   & 0.347 & 480   & 4.6 & 0.321 & 8317  & 153.5  \bigstrut\\
\cline{1-7} PruneFL & 0.355 & 457   & 4.6 & 0.303 & 18817 & 153.5 \bigstrut\\
\cline{1-7} \ThePruneName{} & \textbf{0.361} & \textbf{360} & \textbf{2.5} & \textbf{0.333} & \textbf{7689} & \textbf{59.7} \bigstrut\\
    \hline
    \end{tabular}%
  \label{tab:cnn_vgg_prune_cmp_cifar100}%
\end{table}%

\begin{figure}[!t]
\centering
\subfigure[CNN]{
\includegraphics[width=0.23\linewidth]{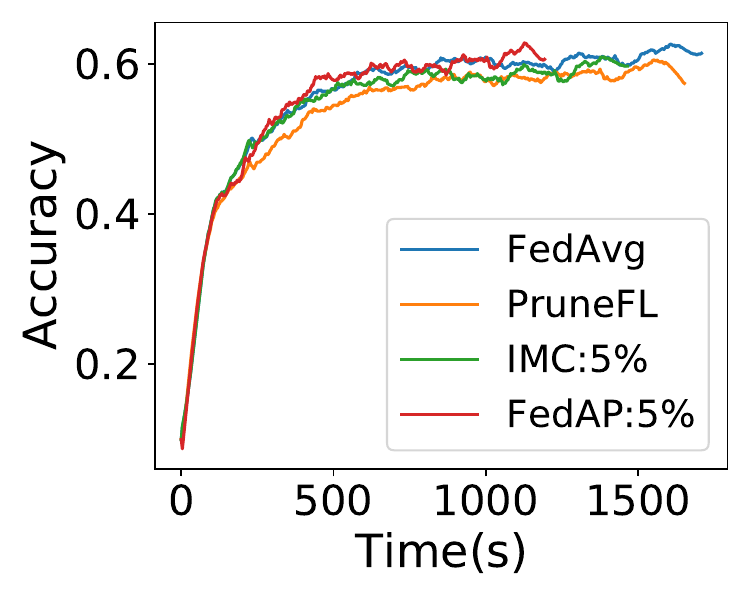}
}
\subfigure[CNN]{
\includegraphics[width=0.23\linewidth]{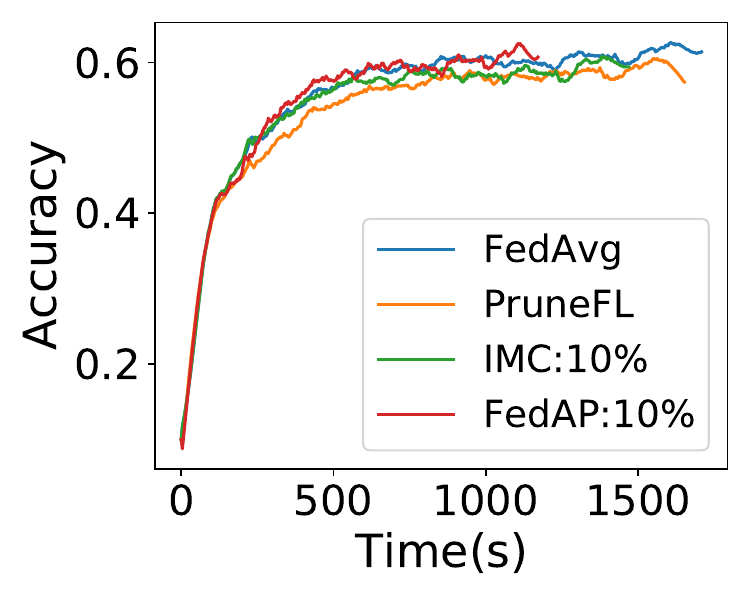}
}
\subfigure[ResNet]{
\includegraphics[width=0.23\linewidth]{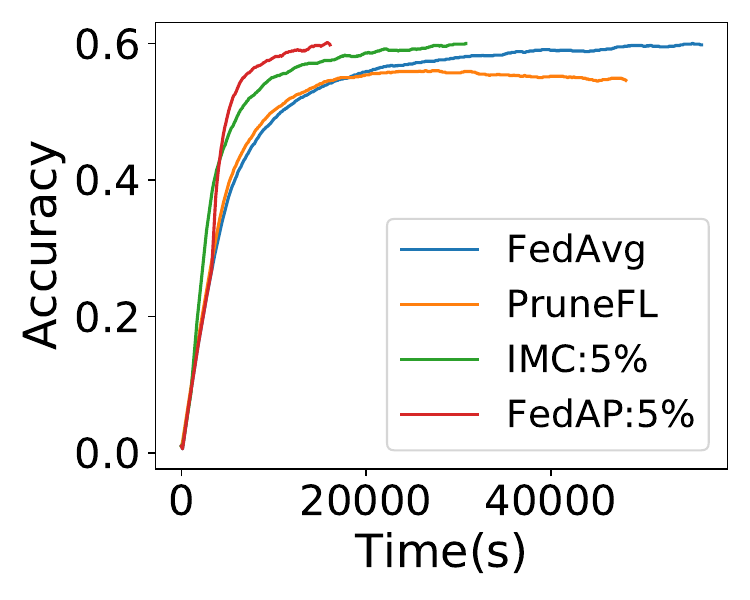}
}
\subfigure[ResNet]{
\includegraphics[width=0.23\linewidth]{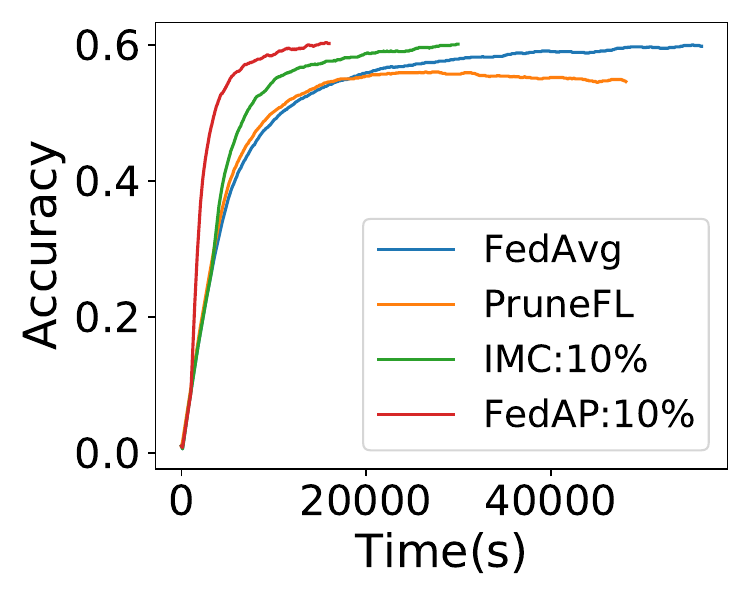}
}
\caption{\textbf{The accuracy and the training time with FedAvg, IMC, PruneFL, and \ThePruneName{} with $p = 5\%$ and $p = 10\%$.} CNN is with CIFAR-10 and ResNet is with CIFAR-100. }
\label{fig:cmp_imc_prunefl}
\end{figure}

\jn{We have similar results on CIFAR-100. As shown in Figure \ref{fig:fedDUM_100}, compared with FedAvg and \TheAlgoName{}, \TheAlgoNameM{} achieves significantly higher accuracy for CNN (up to 1.6\% for FedAvg), VGG (up to 4.2\% for FedAvg, 1.8\% for \TheAlgoName{}), and LeNet (up to 1.1\% for FedAvg, 0.1\% for \TheAlgoName{}), except one case, i.e., slightly lower (0.4\%) for CNN compared with \TheAlgoName{}. The advantages of \TheAlgoNameM{} become significant compared with FedDU-SM (up to 4.4\%), FedDU-DM (up to 1.6\%), and FedDU-DA (up to 6.2\%), in terms of accuracy. In addition, FedDU-DA corresponds to a longer total training time (up to 91.8\% compared with other methods). Furthermore, FedDUM corresponds to a shorter training time to achieve the target accuracy (up to 1.4 times faster than FedAvg, 1.6 times faster than FedDU-SM, 51.8\% times faster than FedDU-DM, and 4.0 times faster than FedDA). }

\subsubsection{Evaluation on \ThePruneName{}}

In this section, we compare \ThePruneName{} with three baseline methods, i.e., HRank \cite{lin2020hrank}, IMC \cite{zhang2021validating}, and PruneFL \cite{jiang2023model}. We utilize HRank with the shared insensitive server data with multiple pruning rates, e.g., $0.2$, $0.4$, $0.6$, and $0.8$, within the FL training process. Similarly, we exploit IMC based on the server data in the FL training process. HRank is a structured pruning method, while PruneFL and IMC are unstructured techniques.



Figure \ref{fig:cmp_prunerate_l5} reveals the results based on CNN and VGG with CIFAR10 in terms of training time. 
\ThePruneName{} can generate a proper pruning rate  (37.8\% on average for CNN and 72.0\% on average for VGG), with a faster training speed (up to 1.8 times faster) and an excellent accuracy (up to 0.1\% accuracy reduction compared with that of FedAvg). However, both the training time and the accuracy decrease at the same time when the pruning rate increases with HRank.
\ThePruneName{} generates adaptive pruning rates for each layer while the accuracy is almost the same as that of FedAvg. 
Then, we compare \ThePruneName{} with FedAvg, IMC, and PruneFL. Figure \ref{fig:cmp_imc_prunefl} demonstrates that the training speed of \ThePruneName{} is significantly higher than baselines. 
In addition, \ThePruneName{} achieves the highest accuracy
while its training time is significantly shorter to achieve the target accuracy of 0.55.
As unstructured pruning techniques cannot reduce computational costs with general-purpose hardware, the training time remains unchanged. On the contrary, \ThePruneName{} corresponds to a much smaller computational cost (up to 55.7\% reduction).

\begin{table}[!t]
  \centering
  \small
  \caption{\textbf{The final accuracy, training time, and computational cost of LeNet with FedAvg, IMC, PruneFL, and \ThePruneName{} based on CIFAR-100.}}
    \begin{tabular}{|c|c|c|c|}
    \hline
    Methods & Accuracy & Time(0.25) &  MFLOPs \bigstrut\\
    \hline
    FedAvg & 0.253 & 897   & 0.7 \bigstrut\\
\cline{1-4} IMC   & 0.257 & 617   & 0.7 \bigstrut\\
\cline{1-4} PruneFL & 0.255 & 804   & 0.7 \bigstrut\\
\cline{1-4} \ThePruneName{} & \textbf{0.259} & \textbf{546} & \textbf{0.6} \bigstrut\\
    \hline
    \end{tabular}%
  \label{tab:lenet_prune_cmp_cifar100}%
\end{table}%

\begin{figure}[!t]
\centering
\subfigure[VGG]{
\includegraphics[width=0.23\linewidth]{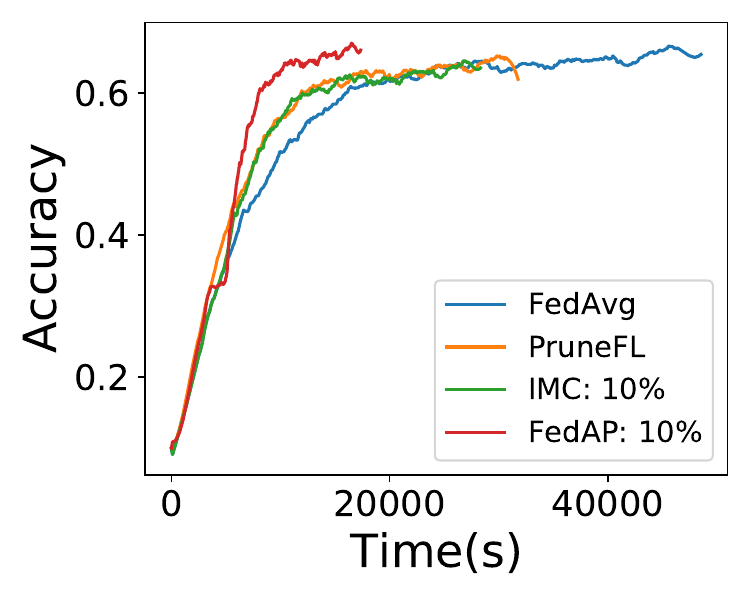}
}
\subfigure[LeNet]{
\includegraphics[width=0.23\linewidth]{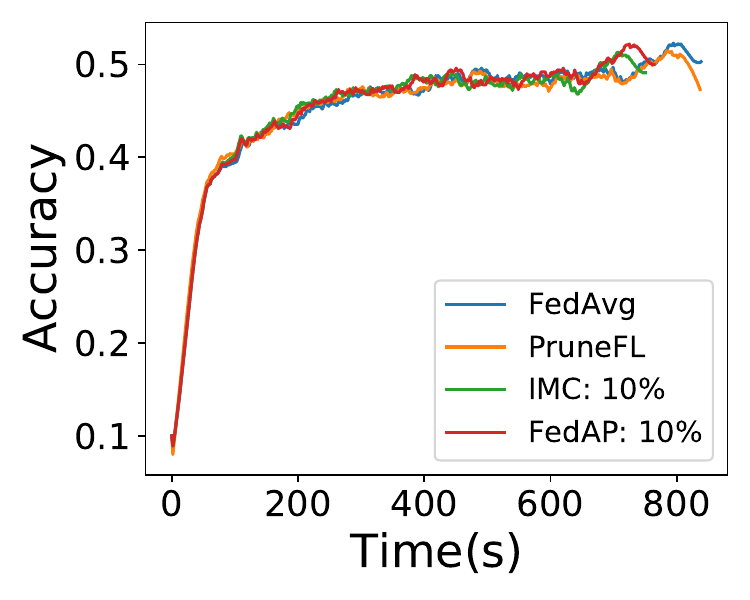}
}
\caption{\textbf{The accuracy and training time with FedAvg, IMC, PruneFL, and \ThePruneName{} based on CIFAR-10.}}
\label{fig:cmp_prune_cifar10}
\end{figure}

\begin{figure}[!t]
\centering
\subfigure[CNN]{
\includegraphics[width=0.23\linewidth]{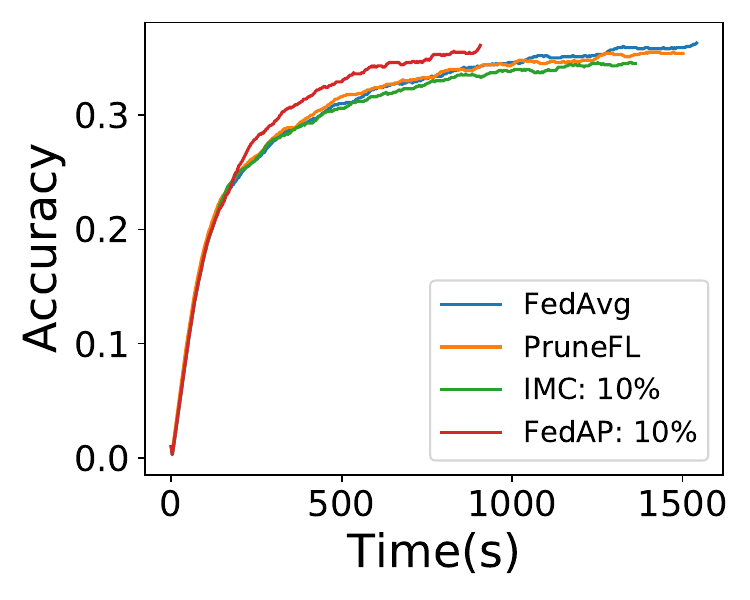}
}
\subfigure[VGG]{
\includegraphics[width=0.23\linewidth]{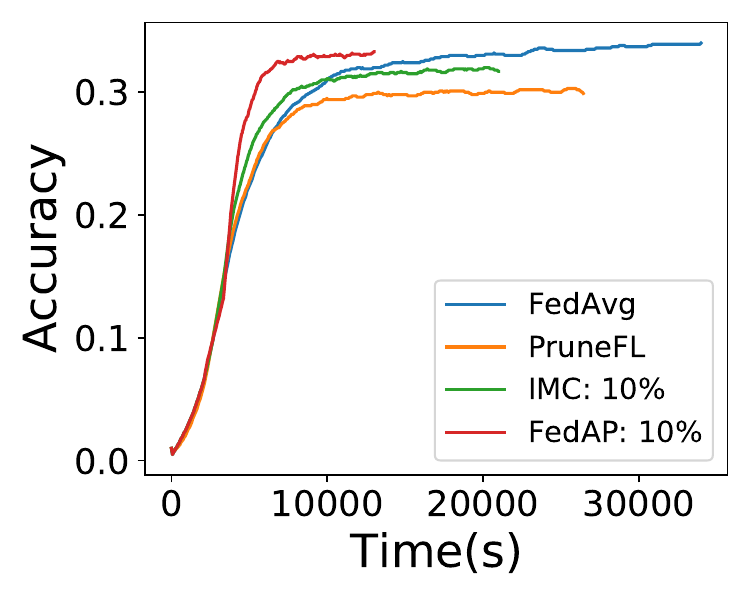}
}
\subfigure[LeNet]{
\includegraphics[width=0.23\linewidth]{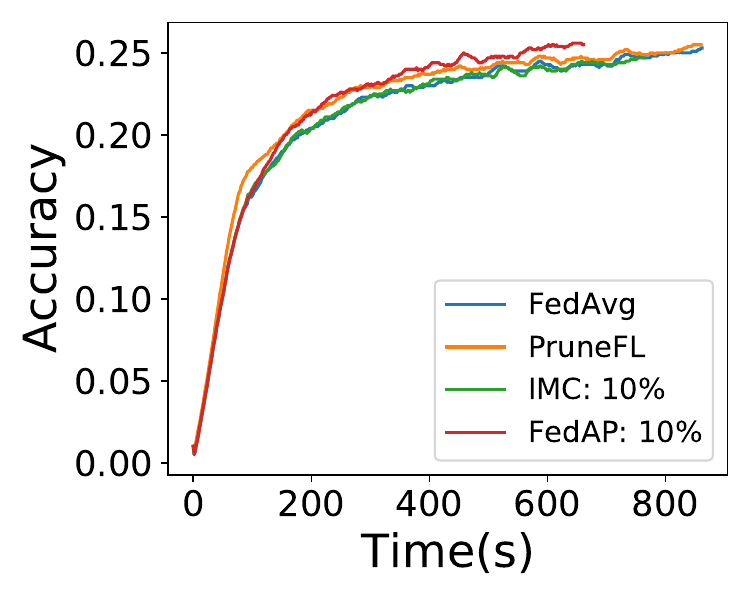}
}
\subfigure[ResNet]{
\includegraphics[width=0.23\linewidth]{fig/resnet/cifar100_prune_cmp_p10_NonIID_Accuracy.pdf}
}
\caption{\textbf{The accuracy and training time with FedAvg, IMC, PruneFL, and \ThePruneName{} based on CIFAR-100.}}
\label{fig:cmp_prune_cifar100}
\end{figure}

\jn{As shown in Figure \ref{fig:cmp_prune_cifar10} and Tables \ref{tab:vgg_prune_cmp_cifar10} and \ref{tab:lenet_prune_cmp_cifar10}, \ThePruneName{} achieves the highest accuracy and incurs a negligible accuracy reduction (up to 0.1\% than FedAvg) while its training time is significantly shorter to achieve the accuracy of 0.6 (up to  one time faster than FedAvg, 50.3\% faster than IMC, and 49.1\% faster than PruneFL). }

\jn{As shown in Figure \ref{fig:cmp_prune_cifar100} and Table \ref{tab:cnn_vgg_prune_cmp_cifar100}, with CNN and VGG on CIFAR-100, \ThePruneName{} achieves the highest accuracy and incurs a negligible accuracy reduction (up to 0.2\% and 0.7\%) while its training time is significantly shorter to achieve the accuracy of 0.3 (up to 8.2\% faster than IMC, 1.4 times faster than PruneFL, and 66.4\% faster than FedAvg). As shown in Table \ref{tab:lenet_prune_cmp_cifar100}, similarly with LeNet, \ThePruneName{} achieves the highest accuracy while its training time is significantly shorter to achieve the accuracy of 0.25 (up to 13.0\% faster than IMC, 47.3\% faster than PruneFL, and 64.3\% faster than FedAvg). }

\subsubsection{\jn{Evaluation on \TheNameM{}}}

\jn{In this section, we compare \TheNameM{}, consisting of both \TheAlgoNameM{} and \ThePruneName{}, with the baseline approaches. We present the comparison results of CNN, VGG, LeNet with CIRAR-10 and CIFAR-100.}

\jn{As shown in Table \ref{tab:finalComparison},
\TheNameM{} with CNN achieves higher accuracy compared with FedAvg (5.8\%), Data-sharing (5.0\%), Hybrid (20.4\%), ServerM (1.4\%), DeviceM (0.8\%), FedDA (2.9\%), IMC (7.6\%), and PruneFL (7.9\%) for CNN. In addition, \TheNameM{} has a shorter training time to achieve the target accuracy (up to 2.4 times faster than FedAvg, 16.9 times faster than Data-sharing, 
4.3 times faster than IMC, and 5.2 times faster than PruneFL) and a much smaller computational cost (up to 41.3\% compared with others).}

\begin{figure}[!t]
\centering
\subfigure[CNN]{
\includegraphics[width=0.23\linewidth]{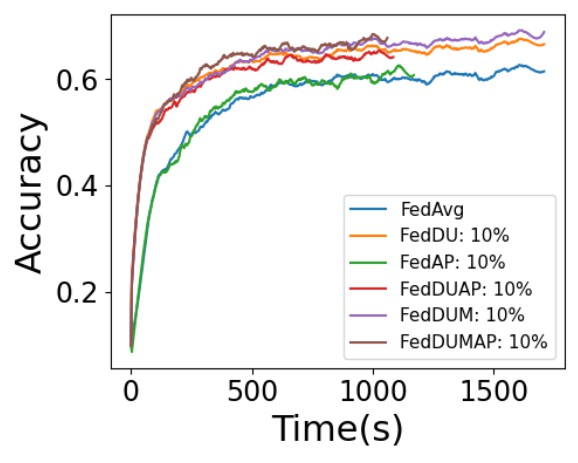}
} 
\subfigure[VGG]{
\includegraphics[width=0.23\linewidth]{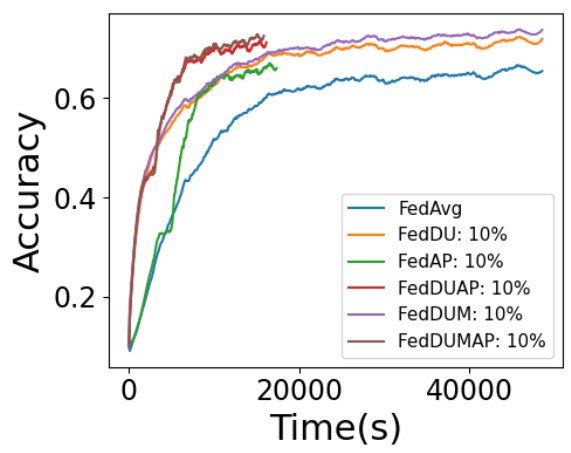}
}
\subfigure[LeNet]{
\includegraphics[width=0.23\linewidth]{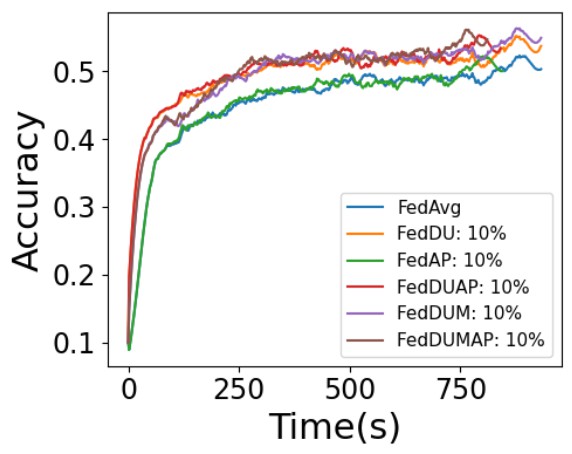}
}
\caption{\textbf{The accuracy of FL with FedAvg, \TheAlgoName{}, \ThePruneName{}, and \TheName{} based on CIFAR-10.}}
\label{fig:ablation}
\end{figure}

\begin{figure}[!t]
\centering
\subfigure[CNN]{
\includegraphics[width=0.23\linewidth]{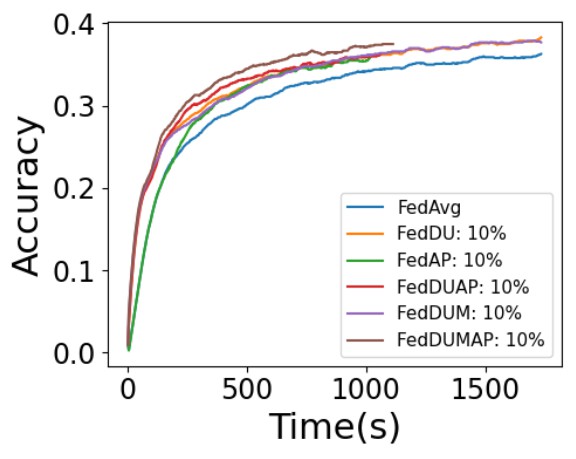}
}  
\subfigure[VGG]{
\includegraphics[width=0.23\linewidth]{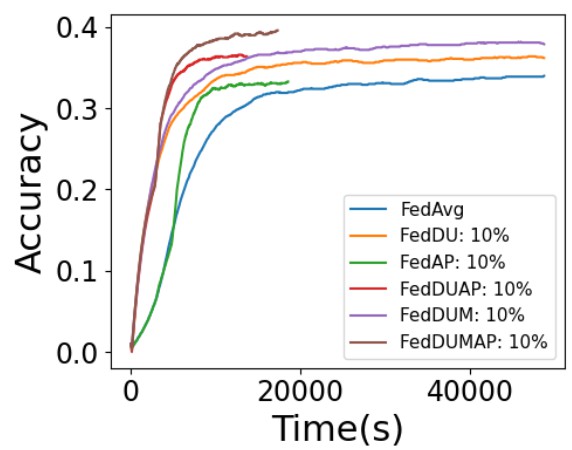}
}
\subfigure[LeNet]{
\includegraphics[width=0.23\linewidth]{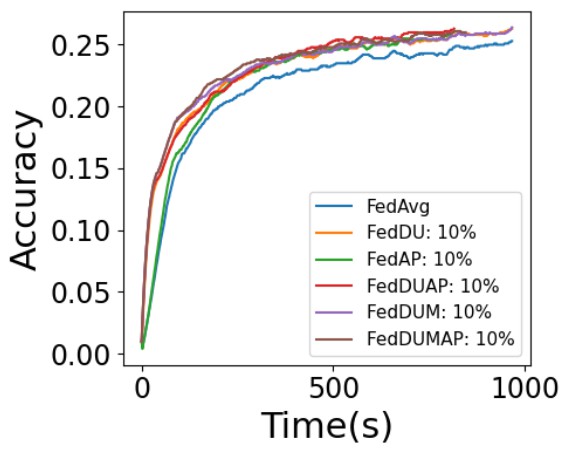}
}
\caption{\textbf{The accuracy of FL with FedAvg, \TheAlgoName{}, \ThePruneName{}, and \TheName{} based on CIFAR-100.}}
\label{fig:ablation_cifar100}
\end{figure}

\begin{table}[!t]
  \centering
  \small
  \caption{\textbf{The final accuracy, training time, and computational cost with diverse approaches on CIFAR-10.} ``Accuracy'' represents the accuracy of the final global model. ``Time'' represents the training time (s) to achieve the accuracy of 0.6 for CNN and VGG, and 0.5 for LeNet. ``MFLOPs'' represents the computational cost of devices. ``NaN'' represents that the accuracy does not achieve the required accuracy. ``D-S'' represents Data-sharing.}
\label{tab:finalComparison}
    \begin{tabular}{|c|c|c|c|c|c|c|c|c|c|}
    \hline
    \multirow{2}[4]{*}{Method} & \multicolumn{3}{c|}{CNN} & \multicolumn{3}{c|}{VGG} & \multicolumn{3}{c|}{LeNet} \bigstrut\\
\cline{2-10} & Accuracy & Time & MFLOPs & Accuracy & Time & MFLOPs & Accuracy & Time & MFLOPs \bigstrut\\
    \hline
    FedAvg & 0.626 & 838   & 4.5 & 0.666 & 15953 & 153.3 & 0.523 & 824   & 0.7 \bigstrut\\
\cline{1-10} D-S   & 0.634 & 4447  & 4.5 & \textbf{0.735} & 6726  & 153.3 & 0.435 & NaN   & 0.7 \bigstrut\\
\cline{1-10} \rev{FedKT}   & 0.458 & 942  & 4.5 & 0.631 & 11772  & 153.3 & 0.442 & 523   & 0.7 \bigstrut\\
\cline{1-10} \rev{FedDF}   & 0.634 & 1047  & 4.5 & 0.547 & 12035  & 153.3 & 0.474 & 513   & 0.7 \bigstrut\\
\cline{1-10} Hybrid-FL & 0.480  & NaN   & 4.5 & 0.656 & 14785 & 153.3 & 0.400   & NaN   & 0.7 \bigstrut\\
\cline{1-10} ServerM   & 0.670 & 383  & 4.5 & 0.694 & 10797  & 153.3 & 0.528 & 346  & 0.7 \bigstrut\\
\cline{1-10} DeviceM & 0.676 & 431  & 4.5 & 0.722 & 10700  & 153.3 & 0.560 & 280  & 0.7 \bigstrut\\
\cline{1-10} FedDA & 0.655 & 582  & 4.5 & 0.695 & 20333  & 153.3 & 0.501 & 1171  & 0.7 \bigstrut\\
\cline{1-10} IMC   & 0.608 & 1316  & 4.5 & 0.645 & 11992 & 153.3 & 0.513 & 733   & 0.7 \bigstrut\\
\cline{1-10} PruneFL & 0.605 & 1536  & 4.5 & 0.652 & 11896 & 153.3 & 0.515 & 837   & 0.7 \bigstrut\\
\cline{1-10} \TheNameM{} & \textbf{0.684} & \textbf{248} & \textbf{2.7} & 0.728 & \textbf{4677} & \textbf{57.3} & \textbf{0.561} & \textbf{249} & \textbf{0.6} \bigstrut\\
    \hline
    \end{tabular}
\end{table}

\begin{table}[!t]
  \centering
  \small
    \caption{\textbf{The final accuracy, training time, and computational cost with diverse approaches based on CIFAR-100.} ``Accuracy'' represents the accuracy of the final global model. ``Time'' represents the training time (s) to achieve the accuracy of 0.3 for CNN and VGG, and 0.25 for LeNet. ``MFLOPs'' represents the computational cost of devices. ``NaN'' represents that the accuracy does not achieve the required accuracy. ``D-S'' represents Data-sharing.}
    \begin{tabular}{|c|c|c|c|c|c|c|c|c|c|}
    \hline
    \multirow{2}[4]{*}{Method} & \multicolumn{3}{c|}{CNN} & \multicolumn{3}{c|}{VGG} & \multicolumn{3}{c|}{LeNet} \bigstrut\\
\cline{2-10} & Accuracy & Time & MFLOPs & Accuracy & Time & MFLOPs & Accuracy & Time & MFLOPs \bigstrut\\
    \hline
    FedAvg & 0.363 & 494   & 4.6 & 0.340  & 12798 & 153.3 & 0.253 & 897   & 0.7 \bigstrut\\
\cline{1-10}  D-S   & 0.243 & NaN   & 4.6 & 0.340  & 11842 & 153.3 & 0.159 & NaN   & 0.7 \bigstrut\\
\cline{1-10} \rev{FedKT}   & 0.223 & 484  & 4.5 & 0.313 & 9770  & 153.3 & 0.170 & 718   & 0.7 \bigstrut\\
\cline{1-10} \rev{FedDF}   & 0.213 & 476  & 4.5 & 0.098 & 11247  & 153.3 & 0.171 & 694   & 0.7 \bigstrut\\
\cline{1-10}  Hybrid-FL & 0.179 & NaN   & 4.6 & 0.308 & 22177 & 153.3 & 0.134 & NaN   & 0.7 \bigstrut\\
\cline{1-10}  ServerM   & 0.342 & 667  & 4.6 & 0.323 & 13775  & 153.3 & 0.256 & 862  & 0.7 \bigstrut\\
\cline{1-10}  DeviceM & \textbf{0.386} & 306  & 4.6 & 0.364 & 8011  & 153.3 & 0.264 & 558  & 0.7 \bigstrut\\
\cline{1-10}  FedDA & 0.349 & 777  & 4.6 & 0.319 & 26422  & 153.3 & 0.258 & 891  & 0.7 \bigstrut\\
\cline{1-10}  IMC   & 0.347 & 480   & 4.6 & 0.321 & 8317  & 153.3 & 0.257 & 617   & 0.7 \bigstrut\\
\cline{1-10}  PruneFL & 0.355 & 457   & 4.6 & 0.303 & 18817 & 153.3 & 0.255 & 804   & 0.7 \bigstrut\\
\cline{1-10}  \TheNameM{} & 0.383 & \textbf{202} & \textbf{2.8} & \textbf{0.396} & \textbf{3849} & \textbf{61.8} & \textbf{0.270} & \textbf{393} & \textbf{0.6} \bigstrut\\
    \hline
    \end{tabular}%
  \label{tab:cnn_vgg_final_cmp_cifar100}%
\end{table}%



\jn{As shown in 
Table \ref{tab:finalComparison}, \TheNameM{} with VGG achieves a higher accuracy compared with FedAvg (6.2\%), IMC (8.3\%), PruneFL (7.6\%), Hybrid-FL (7.2\%), ServerM (3.4\%), DeviceM (0.6\%), FedDA (3.3\%), IMC (8.3\%), and PruneFL (7.6\%). Although slightly outperforming \TheNameM{} (0.7\%), Data-sharing incurs severe data security problems. In addition, \TheNameM{} corresponds to a shorter training time to achieve the accuracy of 0.6 (up to 2.4 times faster than FedAvg, 43.8\% faster than Data-sharing, 2.2 times faster than Hybrid-FL, 1.6 times faster than IMC and 1.5 times faster than PruneFL) and a much smaller computational cost (up to 62.6\% reduction for FedAvg, Data-sharing, Hybrid-FL, IMC, and PruneFL). }


\jn{We obtain similar findings with LeNet. As shown in  Table \ref{tab:finalComparison}, \TheNameM{} achieves a higher accuracy compared with FedAvg (3.8\%), Data-sharing (12.6\%), Hybrid-FL (16.1\%), ServerM (3.3\%), DeviceM (0.1\%), FedDA (6.0\%), IMC (4.8\%) and PruneFL (4.6\%). In addition, \TheNameM{} corresponds to a shorter training time to achieve the accuracy of 0.5 (up to 2.3 times faster than FedAvg, 1.9 times than IMC, 2.4 times than PruneFL) and a much smaller computational cost (up to 14.3\% reduction for FedAvg, Data-sharing, Hybrid-FL, IMC, and PruneFL).}


\jn{As shown in 
Tables \ref{tab:cnn_vgg_final_cmp_cifar100}, 
with CNN, VGG, and LeNet on CIFAR-100, \TheNameM{} achieves a higher accuracy compared with FedAvg (up to 5.6\%), Data-sharing (up to 14.0\%), Hybrid-FL (up to 20.4\%), ServerM (up to 7.3\%), DeviceM (up to 3.2\%), FedDA (up to 7.7\%), IMC (up to 7.5\%) and PruneFL (up to 9.3\%), except for a single case, i.e., slightly lower than DeviceM (0.3\%) for CNN. In addition, \TheNameM{} corresponds to a shorter training time to achieve the accuracy of 0.3 (up to 2.3 times faster than FedAvg, 2.1 times faster than Data-sharing, 4.8 times faster than Hybrid-FL, 1.4 times faster than IMC and 3.9 times faster than PruneFL) and a much smaller computational cost (up to 59.7\% reduction compared with other methods).}



\begin{table}[!t]
  \centering
  \small
  \caption{\textbf{The final accuracy, training time, and computational cost with FedAvg, \TheAlgoName{}, \ThePruneName{}, and \TheName{} with CIFAR-10.} ``Accuracy'' represents the accuracy of the final global model. ``Time'' represents the training time (s) to achieve the accuracy of 0.6 for CNN and VGG, and 0.5 for LeNet. ``MFLOPs'' represents the computational cost of devices.}
    \begin{tabular}{|c|c|c|c|c|c|c|c|c|c|}
    \hline
    \multirow{2}[4]{*}{Method} & \multicolumn{3}{c|}{CNN} & \multicolumn{3}{c|}{VGG} & \multicolumn{3}{c|}{LeNet} \bigstrut\\
\cline{2-10} & Accuracy & Time & MFLOPs & Accuracy & Time & MFLOPs & Accuracy & Time & MFLOPs \bigstrut\\
    \hline
    FedAvg & 0.626 & 838   & 4.5 & 0.666 & 15953 & 153.3 & 0.523 & 824   & 0.7 \bigstrut\\
\cline{1-10} \TheAlgoName{} & 0.675 & 284   & 4.5 & 0.723 & 8268  & 153.3 & 0.552 & 284   & 0.7 \bigstrut\\
\cline{1-10} \TheAlgoNameM{} & \textbf{0.691} & 311 & 4.5 & \textbf{0.738} & 7490  & 153.3 & \textbf{0.563} & 284   & 0.7 \bigstrut\\
\cline{1-10} \ThePruneName{} & 0.625 & 696   & 3.0 & 0.670  & 7980  & \textbf{56.0} & 0.522 & 754   & \textbf{0.6} \bigstrut\\
\cline{1-10} \TheName{} & 0.662 & 278 & 2.8 & 0.714 & 4719 & 58.4 & 0.553 & 267 & \textbf{0.6} \bigstrut\\
\cline{1-10} \TheNameM{} & 0.684 & \textbf{248} & \textbf{2.7} & 0.728 & \textbf{4677} & 57.3 & 0.561 & \textbf{249} & \textbf{0.6} \bigstrut\\
    \hline
    \end{tabular}%
  \label{tab:ablation}%
\end{table}%

\subsubsection{\jn{Ablation Study}}

\jn{We conduct the ablation study by measuring the final accuracy, the training time to achieve the accuracy of 0.6 for CNN and VGG, 0.3 for LeNet with CIFAR-10, and the computational cost of FedAvg, \TheAlgoName{}, \TheAlgoNameM{}, \ThePruneName{}, \TheName{}, and \TheNameM{}. As shown in Figure \ref{fig:ablation} and Table \ref{tab:ablation}, the efficiency of \TheNameM{} significantly outperforms \TheAlgoName{}, \TheAlgoNameM{}, and FedAvg (up to 2.4 times faster), while the accuracy of \TheNameM{} is much higher than that of \TheAlgoName{}, \ThePruneName{}, and FedAvg (up to 6.2\%). 
In addition, the computational cost of \TheNameM{} is the smallest for CNN, while it is slightly higher (2.1\%) than \ThePruneName{} for VGG due to the dynamic server update. Although \TheNameM{} achieves slightly lower accuracy compared with \TheAlgoNameM{} (up to 1.0\%), \TheNameM{} is much more efficient than \TheAlgoName{} and \TheAlgoNameM{} (up to 76.8\% faster). }

\jn{In order to reveal the advantages of \TheNameM{}, we also conduct the ablation study with CIFAR-100. We measure the final accuracy, the training time to achieve the accuracy, 
and the computational cost of diverse methods.}


\jn{As shown in Figure \ref{fig:ablation_cifar100} and Table \ref{tab:abalation_cnn_vgg_cifar100}, the efficiency of \TheNameM{} significantly outperforms \TheAlgoName{} (up to 2.3 times faster), \TheAlgoNameM{} (up to 43.1\% faster), and FedAvg (up to 2.3 times faster), while the accuracy of \TheNameM{} is much higher than that of \TheAlgoName{} (up to 0.2\%), \ThePruneName{} (up to 6.3\%), and FedAvg (up to 5.6\%). \TheNameM{} corresponds to the shortest training time with CNN, VGG, and LeNet. In addition, the computational cost of \TheNameM{} is the smallest for LeNet while it is slightly higher than \ThePruneName{} for CNN (16.0\%) and VGG (3.5\%) due to the dynamic server update. 
Furthermore, \TheNameM{} has slightly higher accuracy (up to 1.4\% for VGG) compared with that of \TheAlgoNameM{}, and the accuracy of \TheNameM{} is much higher (up to 5.6\% for VGG) than that of FedAvg. In addition, \TheNameM{} is much more efficient than \TheAlgoNameM{} (up to 43.1\% faster for CNN). }

\begin{table}[!t]
  \centering
  \small
  \caption{\textbf{The final accuracy, training time, and computational cost with FedAvg, \TheAlgoName{}, \ThePruneName{}, and \TheName{} based on CIFAR-100.} ``Accuracy'' represents the accuracy of the final global model. ``Time'' represents the training time (s) to achieve the accuracy of 0.3 for CNN and VGG, and 0.25 for LeNet. ``MFLOPs'' represents the computational cost of devices.}
    \begin{tabular}{|c|c|c|c|c|c|c|c|c|c|}
    \hline
    \multirow{2}[4]{*}{Method} & \multicolumn{3}{c|}{CNN} & \multicolumn{3}{c|}{VGG} & \multicolumn{3}{c|}{LeNet} \bigstrut\\
\cline{2-10} & Accuracy & Time & MFLOPs & Accuracy & Time & MFLOPs & Accuracy & Time & MFLOPs \bigstrut\\
    \hline
    FedAvg & 0.363 & 494   & 4.6 & 0.340  & 12798 & 153.3 & 0.253 & 897   & 0.7 \bigstrut\\
\cline{1-10} \TheAlgoName{} & 0.373 & 667   & 4.6 & 0.364 & 6155  & 153.3 & 0.263 & 574   & 0.7 \bigstrut\\
\cline{1-10} \TheAlgoNameM{} & 0.379 & 355   & 4.6 & 0.382 & 5276  & 153.3 & 0.264 & 572  & 0.7 \bigstrut\\
\cline{1-10} \ThePruneName{} & 0.361 & 360   & 2.4 & 0.333 & 7689  & \textbf{59.6} & 0.259 & 546   & \textbf{0.6} \bigstrut\\
\cline{1-10} \TheName{} & 0.363 & 269 & 2.6 & 0.366 & 3913 & 62.6 & 0.263 & 485 & \textbf{0.6} \bigstrut\\
\cline{1-10} \TheNameM{} & \textbf{0.383} & \textbf{202} & \textbf{2.8} & \textbf{0.396} & \textbf{3849} & 61.8 & \textbf{0.270} & \textbf{393} & \textbf{0.6} \bigstrut\\
    \hline
    \end{tabular}%
  \label{tab:abalation_cnn_vgg_cifar100}%
\end{table}%

\section{Conclusion}
\label{sec:conclusion}
In this paper, we propose a novel FL framework \TheNameM{}, which leverages both the shared insensitive server data and the distributed sensitive device data to train the global model. Furthermore, the non-IID degrees of the data is also considered in the FL training process. \TheNameM{} consists of a dynamic update FL algorithm, i.e., \TheAlgoName{}, \jn{a simple yet efficient adaptive optimization method on top of \TheAlgoName{}, i.e., \TheAlgoNameM{}, }and an adaptive pruning method, i.e., \ThePruneName{}. 
We conduct extensive experimentation to evaluate the performance of \TheNameM{} with different models and real-life datasets. According to the experimental results, \TheNameM{} significantly outperforms the baseline approaches in terms of accuracy (up to \jn{20.4}\% higher), efficiency (up to \jn{16.9} times faster), and computational cost (up to \jn{62.6}\% smaller).



\bibliographystyle{IEEEtran}     
\bibliography{bibfile}



\end{document}